\documentstyle[aps,floats,psfig]{revtex}

\begin{document}
\psfigurepath{.:plot:figure}
\twocolumn[\hsize\textwidth\columnwidth\hsize\csname @twocolumnfalse\endcsname
\preprint{draft \today}
\title{Magnetic correlations and quantum criticality
in the insulating antiferromagnetic, insulating spin liquid,
renormalized Fermi liquid, and metallic antiferromagnetic
phases of the Mott system V$_2$O$_3$} 
\author{Wei Bao\cite{byline}}
\address{Physics Department, Brookhaven National Laboratory, 
Upton, New York 11973}
\author{C. Broholm}
\address{Department of Physics and Astronomy, 
The Johns Hopkins University, Baltimore, Maryland 21218\\
and Center for Neutron Research,
National Institute of Standards and Technology, 
Gaithersburg, Maryland 20899}
\author{G. Aeppli}
\address{NEC, 4 Independence Way, Princeton, New Jersey 08540}
\author{S. A. Carter}
\address{Department of Physics, University of California, 
Santa Cruz, California 95064}
\author{P. Dai}
\address{Solid State Division, Oak Ridge National Laboratory, 
Oak Ridge, Tennessee 37831}
\author{T. F. Rosenbaum}
\address{James Franck Institute and Department of Physics, 
University of Chicago, Chicago, Illinois 60637}
\author{J. M. Honig and P. Metcalf}
\address{Department of Chemistry, Purdue University, West Lafayette, 
Indiana 47907}
\author{S. F. Trevino}
\address{United States Army Research Laboratory, Adelphi, Maryland 20783\\
and Center for Neutron Research,
National Institute of Standards and Technology, 
Gaithersburg, Maryland 20899}
\date{\today}
\maketitle
\begin{abstract}
Magnetic correlations in all four phases of  pure and doped
vanadium sesquioxide ($\rm V_2O_3$) have been examined by 
magnetic thermal neutron scattering. Specifically, we have studied
the antiferromagnetic and paramagnetic phases of metallic  $\rm V_{2-y}O_3$,
the antiferromagnetic insulating and paramagnetic metallic phases of
stoichiometric $\rm V_2O_3$,
and the antiferromagnetic and paramagnetic phases of insulating
$\rm V_{1.944}Cr_{0.056}O_3$. 
While the antiferromagnetic insulator can be accounted for by a localized 
Heisenberg spin model, the long range order in the antiferromagnetic metal
is an incommensurate spin-density-wave, resulting from a
Fermi surface nesting instability. Spin dynamics in the
strongly correlated metal are dominated by spin fluctuations 
with a 
``single lobe'' spectrum in the Stoner electron-hole continuum. 
Furthermore, our results in metallic
V$_2$O$_3$ represent an unprecedentedly complete characterization
of the spin fluctuations near a metallic quantum critical point,
and provide quantitative support for the self-consistent 
renormalization theory for itinerant antiferromagnets in the small moment
limit.
Dynamic magnetic correlations for $\hbar\omega <k_BT$
in the paramagnetic insulator carry
substantial magnetic spectral weight. However, they are extremely 
short-ranged, extending only to the nearest neighbors. 
The phase transition to the antiferromagnetic insulator,
from  the paramagnetic metal and the paramagnetic insulator, 
introduces a sudden switching of magnetic correlations to
a different spatial periodicity which indicates a sudden change in the 
underlying spin Hamiltonian. 
To describe this phase transition and also the 
unusual short range order in the paramagnetic state, 
it seems necessary to take into account the orbital degrees of freedom 
associated with the degenerate $d$-orbitals at the Fermi level 
in $\rm V_2O_3$. 

\vskip1pc
\noindent PACS number(s): 71.27.+a, 75.40.Gb, 75.30.Fv, 71.30.+h
\end{abstract}

\vskip2pc]

\narrowtext

\section{INTRODUCTION}

The Mott metal-insulator transition\cite{bibnfmi} is a localization
phenomenon driven by  Coulomb repulsion between electrons. Mott
insulators are  common among transition metal oxides and are an
important group of materials which cannot be accounted for by
conventional band theories of solids.  
Using a simplified model which
now bears his name,\cite{bibhubm} Hubbard 
demonstrated a metal-insulator transition, in which a half-filled band
is split by electronic correlations into upper (unfilled) and lower
(filled) Hubbard bands.\cite{bibjhc}  
His approach, however, failed
to produce a Fermi surface on the metallic side of the Mott
transition\cite{bibhubc}. Also, there were unresolved questions concerning 
electronic spectral weight of the lower Hubbard band.\cite{hubrev77}  
Starting from the metallic side, a
Fermi liquid description of the correlated metal was provided by
Brinkman and Rice\cite{bibwfbrb}.  They found that while the Pauli spin
susceptibility, $\chi$, and the Sommerfeld constant, $\gamma$, are both
strongly enhanced by Coulomb repulsion, the Wilson ratio,
$\chi/\gamma$, remains constant.  Such behavior reflects a strongly
enhanced effective mass and it was found that the metal to insulator transition
could be associated with a divergence in the effective mass of these
Fermionic quasi-particles.

The discovery of high temperature superconductivity in cuprates has
revived interest in Mott systems\cite{pwa_mott}. To be precise, the
cuprates are classified as charge transfer systems\cite{zsa_ct}, but
similar low energy physics is expected for both classes of materials.  
Progress in experimental techniques, such as
photo-emission and X-ray absorption spectroscopy, has recently
enabled  direct measurements of the Hubbard bands
and the Brinkman-Rice resonance.\cite{fujia} 
On the theoretical side,
there has also been progress especially in using infinite dimensional
mean field theories\cite{inftymv} to calculate the local spectral
function of the Hubbard model close to the metal-insulator
transition\cite{inftymj}.  A particularly important
result is the synthesis of the seemingly contradictory pictures of
Hubbard, Brinkman-Rice, and Slater on the metal-insulator transition in
strongly correlated electron systems\cite{inftymitb}.  It is
encouraging that these D=$\infty$ theories to a large extent reproduce
the experimentally observed features\cite{fujia} 
in the electronic density of
states for three dimensional (3-D) transition metal oxides.

While there has been  a remarkable convergence of experiments and
theories which probe the electronic density of states across the Mott
metal-insulator transition, little work has been done to explore
magnetism close to this phase boundary for materials apart from
the laminar cuprates. On the insulating side, the
Heisenberg spin Hamiltonian was expected to offer a good description of
magnetism but as we shall see in the following,  this turns out not to be
the case for systems with degenerate atomic orbitals.  
On the metallic side of the
transition, one might have hoped that the local spin picture would
survive because charge carriers in the Brinkman-Rice liquid are heavy and
nearly localized. As we shall see this is also inconsistent
with experiments. It turns out that the mobile quasi-particles 
have profound effects on magnetism in the metal.
Indeed, the magnetism is that which one would associate with a
quantum critical state, whose parameters are consistent with, 
for example, Moriya's self-consistent renormalization
(SCR) theory of metallic spin 
fluctuations\cite{scrmorb1,scrmd,scrmor0,scrrama,scrlnzr,scrmor}.

\begin{figure}[bt]
\centerline{
\psfig{file=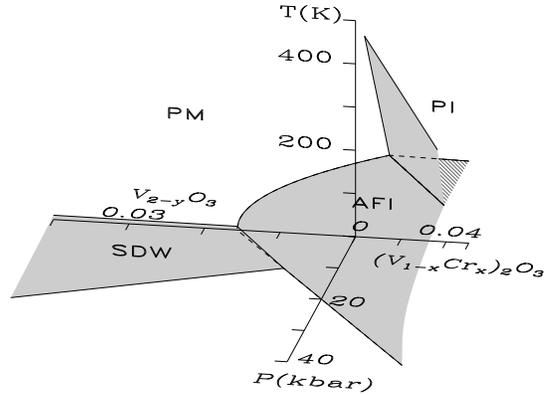,width=\columnwidth,angle=90,clip=}}
\caption{The composition-pressure-temperature phase diagram of the V$_2$O$_3$
system\protect\cite{bibdbml,bibyukkd,bao93}. There exist four phases: the
paramagnetic insulator (PI), the antiferromagnetic insulator (AFI), the
paramagnetic metal (PM), and the spin-density-wave (SDW), a.k.a. 
antiferromagnetic metal (AFM). The Mott transition between 
the PM and PI phases is first order, as well as the transitions 
to the AFI phase. The first order nature 
of the latter phase transition becomes much less pronounced
for Cr  concentrations exceeding $x$$\sim$3\%. The transition
to the SDW is second order. Except for the AFI phase which is
monoclinic, all other phases have  the trigonal corundum structure.}
\label{phs}
\end{figure}

We obtained these surprising conclusions and others to be discussed
below through a neutron scattering study of magnetic correlations in
V$_2$O$_3$ and its doped derivatives which constitute a famous 
3-D Mott system\cite{bibdbml,bibnfmi}.  Even though
magnetic order occurs at low T in the metallic and the insulating
phases of $\rm V_2O_3$, the material is the only known transition metal
oxide to display a Paramagnetic Metal (PM) to Paramagnetic Insulator 
(PI) transition (see Fig.~\ref{phs}).  The PM
to Antiferromagnetic Insulator (AFI) transition in
pure V$_2$O$_3$  is a spectacular first order phase
transition in which the resistivity abruptly increases by  8 orders of
magnitude\cite{bibjf,bibhkjmh}, the lattice structure changes from
trigonal to monoclinic\cite{bibdbmc,bibdbmh}, and the staggered moment
immediately reaches over 80\% of the low temperature
ordered moment\cite{bibrmm}. The mechanism for 
the phase transition remains controversial though it is
clear that electron correlations are important because band structure
calculations predict a metallic state for both the trigonal and
monoclinic crystal structures.\cite{bibinmw0,biblenm} One explanation
for the transition is due to Slater\cite{slateraf} who argued that the
antiferromagnetic transition doubles the unit cell which in turn 
opens a gap at the Fermi energy. 
This mechanism however cannot account for the PM-PI
transition and offers no explanation for the existence of the PI
phase.  Another viewpoint is that the AFI phase is simply the
magnetically ordered state of the PI and that the Mott mechanism works
for the PM-AFI transition as it does for the PM-PI transition.
Following this approach, there have been various attempts to
incorporate the antiferromagnetic phase in the framework of the Mott
metal-insulator transition\cite{mthbcyrl,scrmorv,bibjsad,inftymitb}.  
However, none of the theories convincingly accounts for the
first order nature of the PI-AFI transition observed in 
$\rm V_2O_3$.
Another shortcoming of the theories is that they assume that low energy
and static magnetic correlations are characterized by the same wave
vector throughout the phase diagram, something that our experiments
show is {\em not} the case for the $\rm V_2O_3$ system.

Magnetic order and spin waves below 25 meV in the AFI phase of
V$_2$O$_3$ were previously measured by other
workers\cite{bibrmm,bibrew,bibmysawb}.  Our experiments are the first
neutron scattering study of magnetic correlations in the AFM, PM and PI
phases of V$_2$O$_3$.  Our most important conclusions are the
following:  (1) the AFM has a small moment incommensurate
spin-density-wave (SDW) at low temperatures, which results from a Fermi
surface instability.  (2) Dynamic spin correlations in the entire PM
phase are controlled by the same Fermi surface instability.  (3)
Magnetic correlations in the metallic phases are those one might
expect near a quantum critical point. They are quantitatively 
described by the SCR theory
for itinerant antiferromagnetism in the small moment limit.  (4)
Anomalously short-range dynamic spin correlations in the PI are closely
related to those in the PM but are different from the magnetic order in
the AFI.  (5) Phase transitions to the AFI, from either the PM or the
PI, cause an abrupt switch of the magnetic wave vector, signaling a
sudden change of the exchange constants in the spin Hamiltonian.  All
of (1)-(5) require more than a conventional Heisenberg spin
Hamiltonian for a satisfactory explanation. In metallic samples the
geometry of the Fermi surface directly affects magnetic correlations,
while in insulating samples there are low energy orbital degrees of
freedom which are important for describing the unusual PI phase and the
PI-AFI phase transition.  Some of the above results were previously
published in short reports\cite{bao93,bao94a,bao95d,bao96a,bao96c,bao97a,baoth}. 

The organization of the paper is as follows. 
Section II covers experimental details concerning sample preparation and
neutron scattering instrumentation,
sections III, IV, V and VI describe magnetic correlations in the
AFM, PM, AFI and PI phases respectively, and section VII concludes the paper
with a discussion of the overall interpretation of the data.

\section{EXPERIMENTAL DETAILS}

\begin{figure}[bt]
\centerline{
\psfig{file=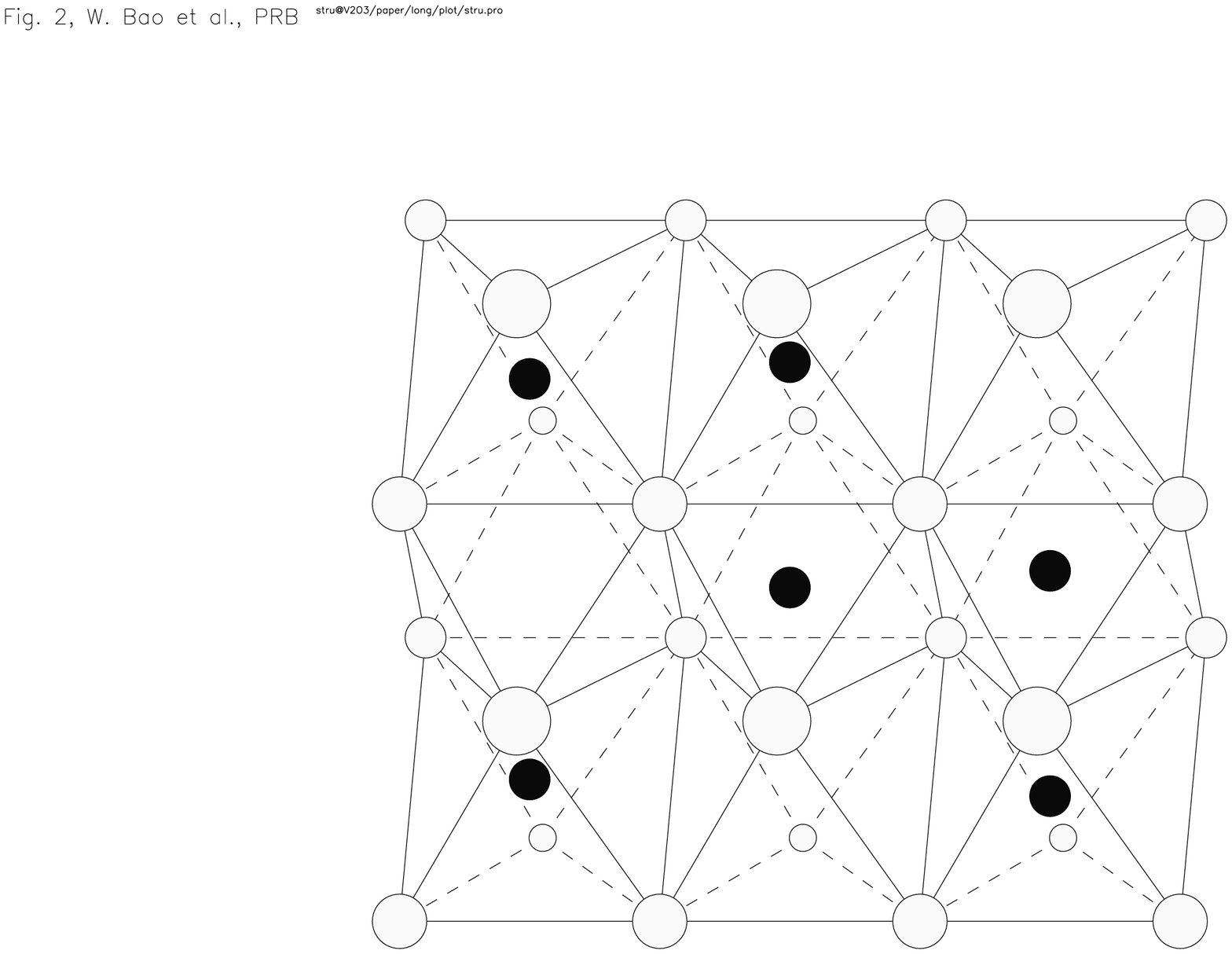,width=\columnwidth,angle=90,clip=}}
\caption{The corundum lattice structure of V$_2$O$_3$.
The solid symbols denote V ions and the hatched symbols O ions.
}
\label{lat_stru}
\end{figure}

\begin{figure}[bt]
\centerline{
\psfig{file=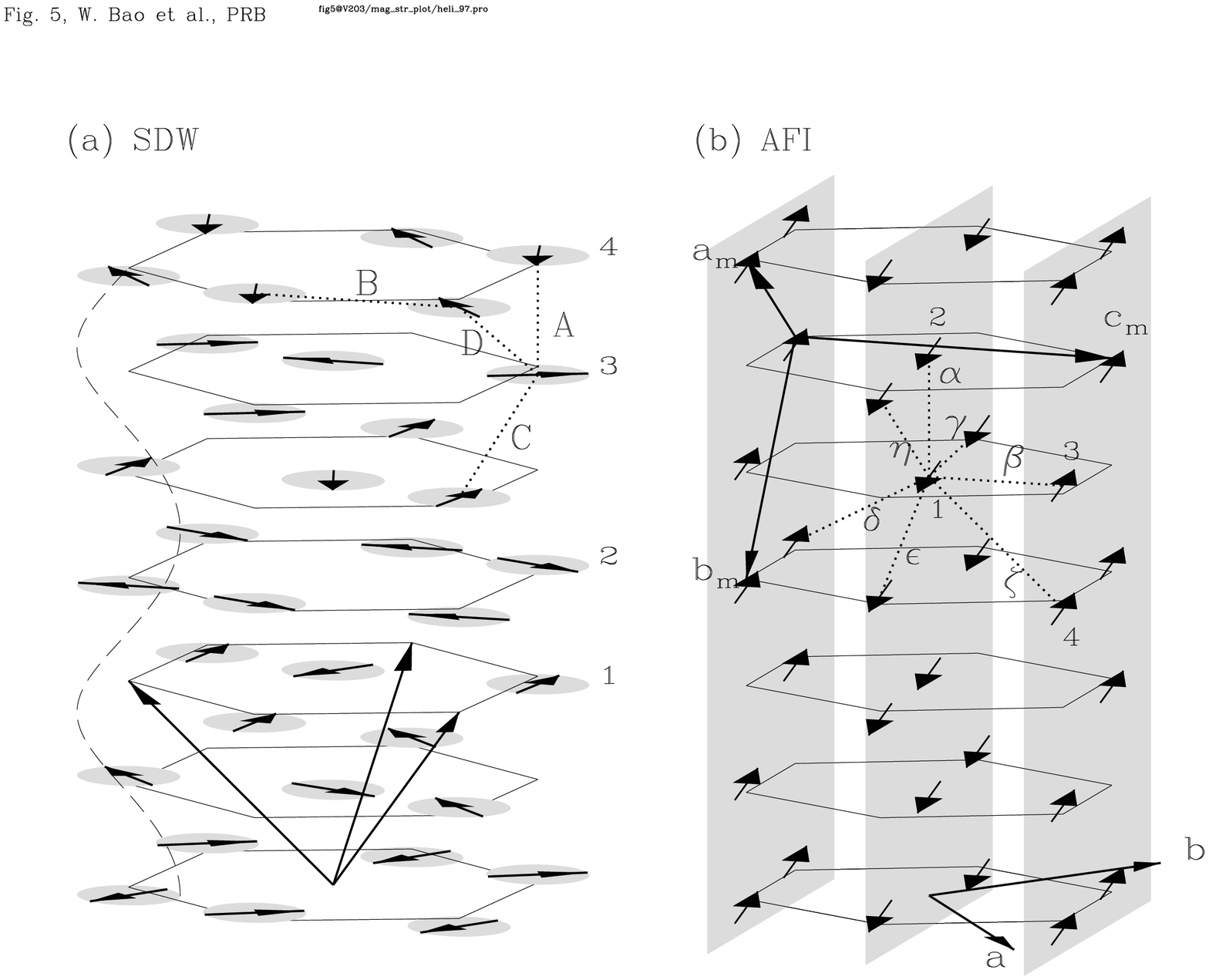,width=\columnwidth,angle=90,clip=}}
\caption{(a) Spin structure in the SDW 
state\protect\cite{bao93} of V$_{2-y}$O$_3$.
Both spiral chiralities exist in our crystal. 
The staggered moment is 0.15$\mu_B$ per V. 
The four types of near neighbor spin pairs 
which are coupled by 
appreciable exchange interactions are indicated by A-D.
(b) Spin structure in the AFI
state\protect\cite{bibrmm}, which breaks the three-fold symmetry of the
corundum structure. The staggered moment is 1.2$\mu_B$ per V. 
$\alpha$-$\eta$ indicate seven
distinct exchange constants.
The conventional hexagonal unit cell is used in this paper.
The primitive vectors {\bf a} and {\bf b} of this cell
are shown in (b), and the {\bf c} vector, which is not shown, spans
the stack from bottom to top. 
The primitive vectors of the {\em primitive} trigonal unit cell are shown 
in (a), and 
the primitive vectors ${\bf a}_m$, ${\bf b}_m$ and ${\bf c}_m$ of the
{\em primitive} monoclinic magnetic unit cell of the AFI
in (b). There are
four spins (1-4) per primitive cell. 
}
\label{spin_stru}
\end{figure}

V$_2$O$_3$ has the corundum structure\cite{bibjbg} 
(space group No.\ 165, $R\overline{3}c$, refer to
Fig.~\ref{lat_stru})
in all but the AFI phase, where small lattice distortions break 
the three fold rotation symmetry to yield a monoclinic structure.\cite{bibdbmh} 
We use the conventional hexagonal unit cell\cite{bao93,bibjf} 
with six V$_2$O$_3$ formula units per unit  cell to index real and reciprocal space.
The reciprocal lattice parameters are $a^* =4\pi/\surd\overline{3}a =
 1.47(1)\AA^{-1}$ and $c^* =2\pi/c = 0.448(1)\AA^{-1}$ in the metallic phases, 
$a^* = 1.46(1)\AA^{-1}$ and $c^* = 0.449(2)\AA^{-1}$ in the AFI phase, 
and $a^* = 1.45(1)\AA^{-1}$ and $c^* = 0.451(2)\AA^{-1}$
in the PI phase. 
More accurate values of lattice parameters, as functions of temperature
and doping, can be found in Ref.~[\cite{bibdbmc,bibwrr,bibwrrsc}].
With this unit cell, the characteristic wave-vector of the
AFI spin structure\cite{bibrmm} is (1/2,1/2,0) while it is
(0,0,1.7) for the SDW structure\cite{bao93} (refer to Fig.~\ref{spin_stru}). 
Samples were oriented in either the ($hhl$) or the ($h0l$) zone to 
reach peaks in the neutron scattering
associated with the two types of magnetic correlations.
Selection rules for nuclear Bragg points in these two zones are
($hhl$): $l=3n$ and ($h0l$): $h-l=3n$, $l=2n$.

\begin{figure}[bt]
\centerline{
\psfig{file=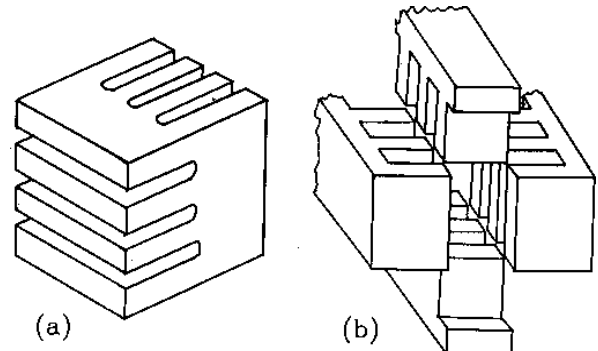,width=.9\columnwidth,angle=-90,clip=}}
\caption{(a) Cross-cut pattern used to facilitate oxygen diffusion
in single crystals during annealing.
(b) Adjustable neutron aperture, made of Boron-polyethylene, to limit
fast neutrons within small angles of the main beam.
Two pairs of opposite ``jaws'' can move into each other
to adjust the central opening according to the sample size.
}
\label{crscut}
\end{figure}

	Single crystals of V$_{2-y}$O$_3$ were grown using a 
skull melter\cite{bibsmpl}. 
The stoichiometry of pure V$_2$O$_3$ and metal-deficient V$_{2-y}$O$_3$
was controlled to within $\delta y=0.003$ by annealing sliced, 
as-grown V$_2$O$_3$ crystals in a suitably chosen CO-CO$_2$ 
atmosphere\cite{bibsmplb} for two weeks at 1400$^o$C. 
Single crystals of (V$_{1-x}$Cr$_{x}$)$_2$O$_3$ were grown using the
Tri-arc technique\cite{bibhkjmh}. The compositions were determined
by atomic absorption spectroscopy and wet chemical titration technique.
The mass of samples for elastic neutron scattering was typically 100 mg, 
while we used several grams for inelastic neutron scattering experiments. 
Large single crystals
were sliced into plates of $\sim$1.5 mm thickness 
as shown in Fig.~\ref{crscut}(a) to allow oxygen 
diffusion to the bulk during the annealing process. 
Several single crystals of each
composition were mutually aligned to increase the sensitivity for
inelastic measurements. We report experiments for the seven compositions 
which are listed in Table~\ref{tbl_sml}. The samples 
\begin{table}
\caption{Samples used in this study. For easy location of the samples in the 
phase diagram (Fig.~\ref{phs}), $x$ or $y$ in molecular formula 
(V$_{1-x}$Cr$_x$)$_2$O$_3$ or V$_{2-y}$O$_3$ is provided. The ambient
transition temperature and physical states below and above the phase
transition also are listed.  }
\label{tbl_sml}
\begin{tabular}{lddcc} 
& $y$ or $x$ & $T_N$ (K) & 0 K & 300 K\\
\hline
V$_{1.963}$O$_3$ & 0.037 & 8.4(1) & SDW & PM \\
V$_{1.973}$O$_3$ & 0.027 & 8.5(1) & SDW & PM \\
V$_{1.983}$O$_3$ & 0.017 & 9.5(1) & SDW & PM \\
V$_{1.985}$O$_3$ & 0.015 & 55(1) & AFI & PM \\
V$_{1.988}$O$_3$ & 0.012 & 70(2) & AFI & PM \\
V$_{2}$O$_3$ & 0.0 & 170(2) & AFI & PM \\
V$_{1.944}$Cr$_{0.056}$O$_3$ & 0.028 & 180(2) & AFI & PI \\
\end{tabular}
\end{table}
have been well characterized in previous studies using bulk measurement
techniques\cite{bibhkjmh,bibsas,bibsacc,bibsctfr,bibsac,bibsacd}.

The sample temperature was controlled using liquid He flow cryostats
in the range from 1.4~K to 293~K, or using displex closed cycle 
cryostats from 10~K to 293~K. With a heating element, the displex 
cryostat could reach temperatures up to 600~K.

	Hydrostatic pressure up to 8 kbar was produced using a He gas 
high pressure apparatus\cite{bao94b} at NIST, which can be loaded into
an ``orange'' ILL cryostat with a temperature range from 1.4~K 
to room temperature.

\subsection{Neutron Scattering Instrumentation}

	Neutron scattering measurements were performed on triple-axis
spectrometers BT2, BT4 and BT9 at the NIST research reactor,
on HB1 at the High Flux Isotope Reactor (HFIR) of ORNL, and
on H4M and H7 at the High Flux Beam Reactor (HFBR) of BNL. 
For unpolarized neutron experiments, we used a
Be (101) monochromator on HB1, Cu (220) and pyrolytic graphite (PG) (002)
monochromators on BT4 and PG (002) monochromators on all other instruments. 
PG (002) analyzers were used for all unpolarized triple-axis experiments. 
For polarized neutron measurements on BT2, crystals of Heusler alloy 
Cu$_2$MnAl [$d$(111)=3.445$\AA$]
were used as monochromator and analyzer. 
Fast neutrons were removed from the incident beam by an in-pile
sapphire filter\cite{saphire} on HB1, and PG filters\cite{pgfilter} 
were used to remove 
high-order neutron contaminations where appropriate.  
The horizontal collimations were controlled with Soller slits 
and are specified in the figures containing the experimental data. 
The fast neutron background at BT4 in the small scattering
angle limit was reduced by a 30 cm long boron-polyethylene 
aperture before the sample and a similar 15 cm long aperture after the sample.
The opening of the aperture was adjusted to match the 
sample size (Fig.~\ref{crscut}(b)). 
These apertures suppress fast neutron background in the small angle limit
which is important for measuring high energy inelastic magnetic scattering. 

\begin{figure}[bt]
\centerline{
\psfig{file=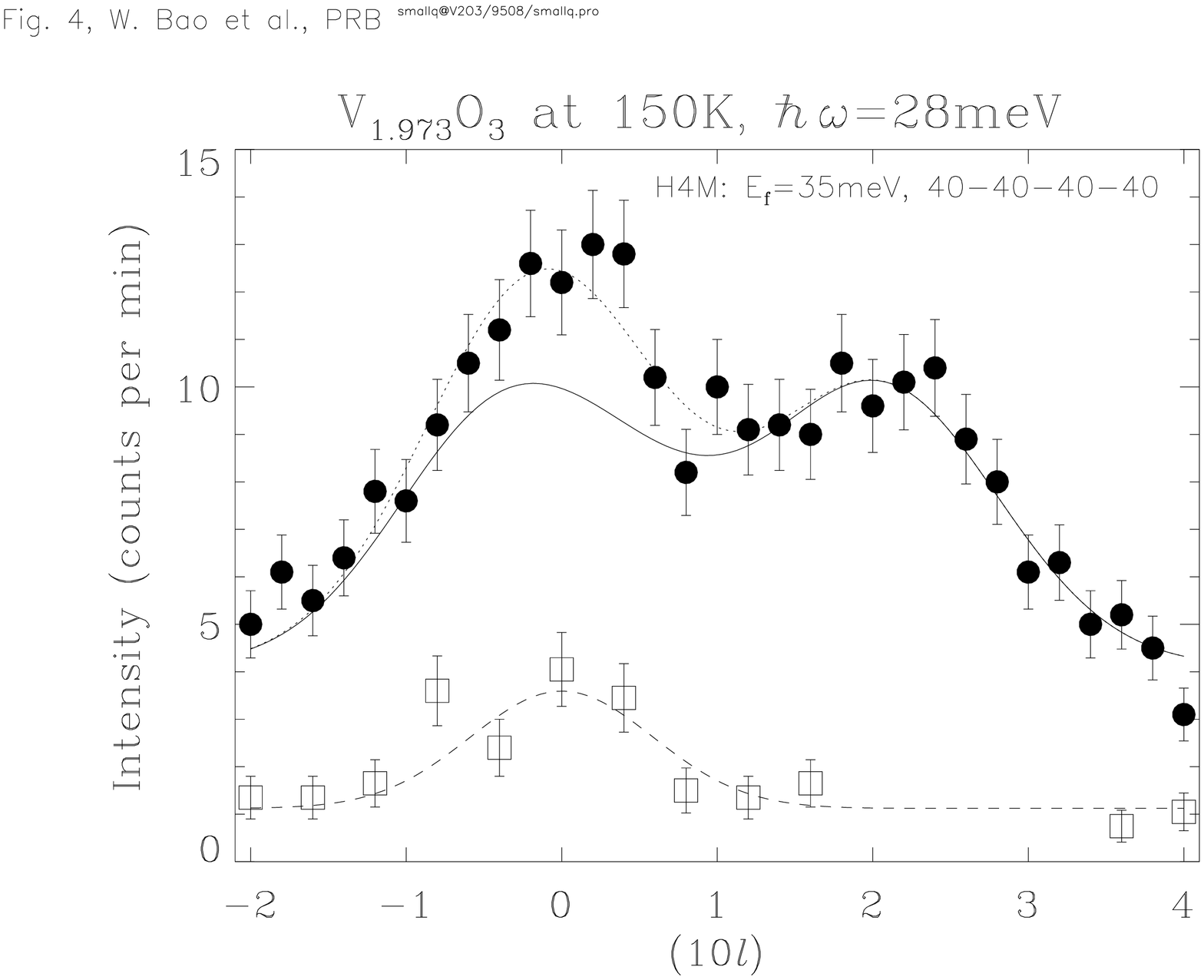,width=\columnwidth,angle=90,clip=}}
\caption{A constant energy scan (solid circles)
for which the scattering angle
near (100) is close to the incident neutron beam.
The squares denote the background measured with 
the analyzer rotated 10$^o$
from the reflection condition.
See text for details.
}
\label{fg_smlq}
\end{figure}

In most of our scans, the background is approximately constant.
For some scans, however, the scattering angle falls below 
$\approx 7^o$ where background increases rapidly with decreasing angle.
For example, a constant-$\hbar\omega=28$meV scan
measured with H4M is shown in Fig.~\ref{fg_smlq} with solid circles. 
The scattering angle at (100) is 5.4$^o$. As a result, the
background (open symbols), measured with the analyzer 
turned 10$^o$ from the reflection condition, increases near (100). 
The solid line represents data with this extra background contribution
subtracted\cite{nt_smlq}.

\subsection{Neutron Scattering Cross Section}

The magnetic neutron scattering cross-section for momentum transfer
$\hbar{\bf q}=\hbar({\bf k}-{\bf k}')$ and energy transfer $\hbar \omega$
is given by \cite{neut_thry}
\begin{eqnarray}
\frac{d^2\sigma}{d\Omega dE'}&=&\frac{k'}{k}
	\left(\frac{\gamma r_0}{2}\right)^2
	\left[f({\bf q})\right]^2e^{-2W({\bf q})}\nonumber \\
&\times &\sum_{\alpha \beta}(\delta_{\alpha\beta}
	-\hat{q}_{\alpha}\hat{q}_{\beta}) 
N{\cal S}^{\alpha\beta}({\bf q},\omega),
\label{cross_sec}
\end{eqnarray}
where $N$ is the number of hexagonal unit cells, $(\gamma r_0/2)^2= 72.65\times 10^{-3}$ barn/$\mu_B^2$, 
$f({\bf q})$ is the magnetic form factor for the 
V$^{3+}$ ion\cite{bibrewajf},
$\exp[-2W({\bf q})]$ is the Debye-Waller factor, 
and the dynamical structure factor is given by
\begin{eqnarray}
{\cal S}^{\alpha\beta}({\bf q},\omega)&=&\frac{(g \mu_B)^2}{2\pi\hbar}
	\frac{1}{N}\sum_{{\bf R\, R}'} \int dt\, e^{i[\omega t-{\bf q}
	\cdot({\bf R}-{\bf R}')]}\nonumber \\
&&\times \langle S^{\alpha}_{\bf R}(t) S^{\beta}_{{\bf R}'}(0)\rangle.
\label{sqw}	
\end{eqnarray}
$g$ is the Land\'{e} factor,
$S^{\alpha}_{\bf R}(t)$ is the
$\alpha$th Cartesian spin component of the V ion at
position {\bf R} at time $t$, and $\langle \ldots \rangle$
denotes a thermal average. 
When spin space anisotropy can be neglected, 
as in the paramagnetic phase of a Heisenberg system, 
expression (\ref{cross_sec}) simplifies to
\begin{equation}
\frac{d^2\sigma}{d\Omega dE'}=2\frac{k'}{k}
	\left(\frac{\gamma r_0}{2}\right)^2
	\left[f({\bf q})\right]^2
	e^{-2W({\bf q})} N{\cal S}^{\alpha\alpha}({\bf q},\omega).
\end{equation}
We will drop the superscripts in the remaining text for brevity. The
structure factor is related to the imaginary part of the dynamic spin
susceptibility by the fluctuation-dissipation theorem
\begin{equation}
{\cal S}({\bf q},\omega)=\frac{1}{1-e^{-\hbar\omega/k_B T}}
	\frac{1}{\pi}\chi''({\bf q},\omega).
\end{equation}
The real part of the generalized spin susceptibility is given by the
Kramers-Kronig relation,
\begin{equation}
\chi'({\bf q},\omega)-\chi'({\bf q},\infty)=\frac{1}{\pi}
\int_{-\infty}^{\infty} d\omega'
\frac{\chi''({\bf q},\omega')}{\omega'-\omega}.
\end{equation}

The intensity in a neutron scattering experiment is measured as
the ratio of neutron counts in a high efficiency $^3$He detector 
to the neutron counts in a
low efficiency monitor placed in the incident beam before the sample.  
To cover a wide
energy range of spin excitations, several spectrometer configurations
were used. The relative sensitivity of different configurations
was determined by comparing the integrated intensities of
identical scans probing magnetic or incoherent nuclear scattering.
Normalizing to inelastic scattering 
from acoustic phonons\cite{bibnjc,MnSi_ishc} yielded absolute measurements 
of ${\cal S}({\bf q},\omega)$ [Eq.~(\ref{sqw})] in units of $\mu_B^2$/meV. 
Units of mbarn/meV are also used in this paper and when we do so 
we are quoting numbers for the following normalized intensity 
\begin{equation}
\tilde{I}({\bf q},\omega )= 2\left(\frac{\gamma r_0}{2}\right)^2
[f({\bf q})]^2{\cal S}({\bf q},\omega),
\end{equation}
which is more directly related to the raw data because it contains 
the magnetic form factor, $f({\bf q})$, as a factor.

\subsection{Resolution Effects}

The measured intensity in a neutron scattering experiment 
is proportional to the convolution 
of the scattering cross-section with an instrumental resolution function.
The resolution function can be approximated
by a Gaussian function in 4-dimensional {\bf q}-$\omega$
space\cite{bibmjc,bibmjce}.
Inelastic neutron scattering experiments are usually flux-limited.
Therefore the choice of an experimental configuration for probing 
a specific scattering cross section always involves a trade-off 
between resolution and sensitivity. The 
following two limits are of importance in this study.

\subsubsection{Infinite life time excitations}

Examples of this type of excitations, such as phonons and un-damped antiferromagnetic spin waves, are described by a scattering cross
section of the form
\[
{\cal S}({\bf q},\omega)\sim\delta(\omega-c|{\bf q}|).\nonumber
\]
If the curvature of the dispersion surface is negligible over the 
volume of the resolution function then
the dispersion surface can be approximated by a
flat plane and the resulting profile of a scan through the dispersion
surface is approximately Gaussian  with
a width defined by the resolution. With coarser resolution, 
the profile  becomes asymmetric and the peak position may
shift from the  position defined by the dispersion relation. 
Refer to the dotted lines in Fig.~\ref{afm_res} for
a few examples of such effects.

\subsubsection{Weakly $\bf q$ and $\omega$ dependent scattering cross section}

Features in the scattering cross-section which are broader in $\bf q$-$\omega$
space than the experimental resolution are reproduced directly
or with slight
additional broadening
in the measured $\bf q$ and $\omega$ dependent intensity. 
This is the case for our measurements
in the AFM, PM and PI phases. 
For example
if ${\cal S}({\bf q},\omega)$ takes the form
\[
{\cal S}({\bf q},\omega)\sim
\exp(-|{\bf q}-{\bf q}_0|^2/\sigma_0^2),
\]
a constant energy scan will yield a Gaussian 
peak with width $(\sigma_0^2+\sigma_R^2)^{1/2}$, 
where $\sigma_R$ is the resolution width. 
To appreciate the resolution broadening, 
let us put in numbers: When the resolution width 
is 1/3 of the intrinsic width,
the measured width  increases by less than 10\% over the intrinsic width.
When the resolution width is 1/5 of the intrinsic width,
the measured width is only 2\% larger than the intrinsic width. 
In this latter case the measured intensity is then very close to being directly proportional 
to ${\cal S}({\bf q},\omega)$.

\section{SDW in metallic V$_2$O$_3$}

The insulating antiferromagnetic state (AFI) of V$_2$O$_3$ can be 
suppressed by a pressure of $\sim$20 kbar\cite{bibdbma,bibsacd}. 
The pressure-stabilized metallic state is paramagnetic down 
to at least 0.35~K.\cite{bibdbmf,bibsacd}
The AFI state can also be suppressed at ambient pressure by vanadium
deficiency\cite{bibmnsh} or titanium doping\cite{bibdbml}. 
The critical concentrations are $x_c \simeq 0.1$ for 
V$_{2-x}$Ti$_x$O$_3$\cite{bibdbml}
and $y_c \simeq 0.015$ for V$_{2-y}$O$_3$\cite{bibsas}.
The existence of antiferromagnetic order at low temperatures 
for the doping induced
metal was first established through the $^{57}$Fe M\"{o}ssbauer 
effect\cite{bibyukk,bibjdcs,bibyukkc}.
The magnetic structure was only recently
determined through single crystal
neutron diffraction\cite{bao93}, 
and it is depicted in Fig.~\ref{spin_stru} (a).
This magnetic order can be described as puckered 
antiferromagnetic honeycomb spin layers
whose spin directions form a helix along the $c$ axis.
The spin directions determined from our neutron diffraction refinement,
which lie in the basal plane,
are consistent with previous susceptibility anisotropy
and magneto-torque
measurements\cite{bibjdcs,bibyukkd}. The pitch of the helix
yields an incommensurate magnetic structure, with
magnetic wave vector 1.7{\bf c}$^*$. Both the staggered moment 
and the wave vector depend weakly on 
doping, $y$.\cite{bao93}  This magnetic order in metallic $\rm V_{2-y}O_3$
bears  little
resemblance to the  magnetic order in the insulating phase of 
$\rm V_2O_3$ 
which is characterized by a magnetic wave vector (1/2,1/2,0)\cite{bibrmm}
[refer to Fig.~\ref{spin_stru} (b)].

We have previously shown\cite{bao93} that 
the antiferromagnetic order in the metallic
state can not be described by a localized spin model,
since the dynamic magnetic spectral weight greatly
exceeds the square of the staggered
moment and the bandwidth for magnetic excitations exceeds k$_B$T$_N$ 
by more than one order of magnitude.
The proper concept is probably that of a 
spin density wave\cite{sdw_lom,sdw_awo,sdw_paf} 
where magnetic order results from a Fermi surface nesting instability.
This interpretation is supported by recent RPA calculation of the
{\bf q}-dependent susceptibility using realistic band structure\cite{sdw_wga}.
According to neutron, specific heat, and transport measurements,
only a small fraction of a large Fermi surface\cite{bao93} is
involved in the SDW of metallic V$_{2-y}$O$_3$.
As seen at the SDW transition for Cr metal\cite{Cr_dbmtmr}, 
the resistivity  for metallic V$_{2-y}$O$_3$
{\em increases} below the N\'{e}el temperature as the Fermi surface
is partially gapped.\cite{bao93} This is in contrast
to localized spin systems where the resistivity {\em decreases} below $T_N$
because of the decrease in spin disorder scattering.
The discovery of the transverse incommensurate 
SDW resolved the longstanding mystery\cite{bibyukkb} about the 
magnetic ground state of doped metallic V$_2$O$_3$.

\subsection{Polarized neutron scattering}

\begin{figure}[bt]
\centerline{
\psfig{file=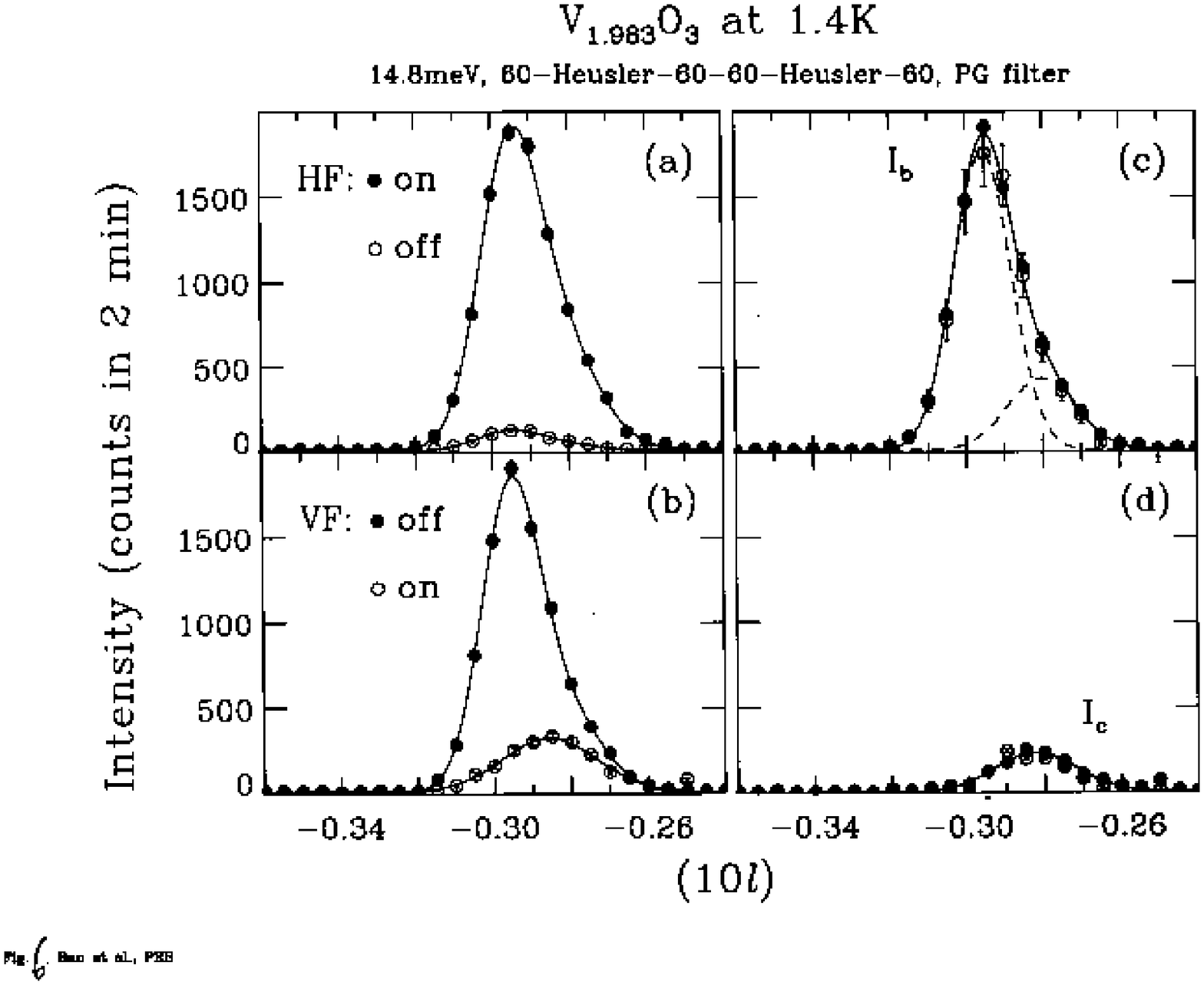,width=\columnwidth,angle=-90,clip=}}
\caption{Triple-axis polarized neutron scans through the 
(1,0,$\overline{0.3}$) magnetic Bragg peak
with neutron spins polarized along the momentum transfer (HF) in (a),
and along the vertical direction (VF) in (b). Measurements with
the neutron spin flipper {\em on} and {\em off} are presented.
The contribution to scattering from spin components
in the basal plane is shown in (c) and the contribution from components
polarized along the {\bf c} axis is shown in (d). See text for details.
}
\label{polar1}
\end{figure}

To further characterize the incommensurate magnetic structure, 
polarized neutron measurements were performed. A V$_{1.983}$O$_3$
sample was oriented with its ($h$0$l$) zone in the horizontal scattering
plane. A small magnetic field of 7.6 G was applied along
either the scattering wave vector (HF) or along the vertical direction (VF)
to guide the neutron spin. A neutron spin flipper was inserted in the neutron
path. In the HF case when the spin flipper is turned {\em on},
only magnetic scattering contributes to the Bragg peak\cite{moonr}.
Coherent nuclear scattering is non-spin-flip and contributes 
when the flipper is
{\em off}. Therefore, Fig.~\ref{polar1}(a) shows that the (1,0,$\overline{0.3}$)
Bragg peak is indeed magnetic. The finite intensity in the flipper off case
is due to incomplete polarization of the neutron beam and is consistent
with a flipping ratio of $f_H$=14, measured at nuclear Bragg peaks.

Since $0.3c^*/a^*\approx 0.09$, 
(1,0,$\overline{0.3}$) is essentially parallel to 
the {\bf a}$^*$ axis. 
The four partial intensities at (1,0,$\overline{0.3}$)
for the HF and VF cases with 
the spin flipper {\em on} or {\em off} are therefore\cite{moonr} 
\begin{mathletters}
\begin{eqnarray}
I^{on}_{HF}&=&I_b+I_c,\label{eq_p1}\\
I^{off}_{HF}&=&(I_b+I_c)/f_H,\\
I^{on}_{VF}&=&I_c+I_b/f_V,\\
I^{off}_{VF}&=&I_b+I_c/f_V,
\label{eq_p4}
\end{eqnarray} \end{mathletters}
where $I_{b(c)} \propto S_{b(c)}^2$ is the contribution from the spin 
components in the ${\bf b} \equiv {\bf a}^* \times {\bf c}^*$ 
 (or {\bf c}) direction, and $f_{H(V)}\simeq 14$ is the
flipping ratio in the HF (VF) case.
Therefore, from the data shown in Fig.~\ref{polar1} (a) and (b), 
$I_b$ and $I_c$ can be extracted by solving the over-determined
equations (\ref{eq_p1})-(\ref{eq_p4}). For example,
\begin{mathletters}
\begin{eqnarray}
I_b&=&I^{off}_{VF}-I_c/f_V \nonumber \\
&\approx &I^{off}_{VF},\\
I_c&=&\frac{I^{on}_{VF}-I^{off}_{HF}}{1-1/f_V}\nonumber \\
&\approx &I^{on}_{VF}-I^{off}_{HF}.
\end{eqnarray} 
\label{eq_p4a}
\end{mathletters}
In the last step in (\ref{eq_p4a}), a small quantity
$I_c/f_V$ is neglected. Alternatively,
\begin{mathletters}
\begin{eqnarray}
I_b&=&I^{on}_{HF}+I^{off}_{HF}-I^{on}_{VF}-I_c/f_V \nonumber\\
&\approx&I^{on}_{HF}+I^{off}_{HF}-I^{on}_{VF},\\
I_c&=&\frac{I^{on}_{HF}-I^{off}_{VF}}{1-1/f_V}\nonumber \\
&\approx &I^{on}_{HF}-I^{off}_{VF}.
\end{eqnarray} 
\label{eq_p4b}
\end{mathletters}
Fig.~\ref{polar1} (c) and (d) show the basal plane component $I_b$ 
and the vertical component $I_c$ derived from the 
data in Fig.~\ref{polar1} (a) and (b) using these expressions. The 
solid circles were determined using Eq.~(\ref{eq_p4a}) and the open circles
using Eq.~(\ref{eq_p4b}).  The two data sets  are  consistent within 
experimental uncertainty and show that the incommensurate peak is 
associated with basal plane spin correlations as was previously
conjectured on the basis of unpolarized neutron diffraction. The data
also reveal a second incommensurate peak with mixed polarization. 
This peak has only been observed in this sample which is very close 
to the critical concentration, $y_c\approx 0.015$, for the
ambient pressure metallic phase \cite{bao93}.

When polarized neutrons are scattered by a right-handed 
transverse spiral in the HF configuration, the partial cross-sections
for magnetic Bragg peaks with wave vectors along the spiral axis are\cite{moonr}
\begin{eqnarray}
\frac{d\sigma({\bf q})}{d\Omega}^{+\rightarrow -}&\propto &
\sum_{{\bf R R}'}e^{i({\bf q}+{\bf q}_m)\cdot({\bf R}-{\bf R}')}\nonumber \\
&\sim&\sum_{\bbox \tau}\delta({\bf q}+{\bf q}_m-{\bbox \tau})	
\end{eqnarray}
for the neutron spin up to spin down channel and
\begin{eqnarray}
\frac{d\sigma({\bf q})}{d\Omega}^{-\rightarrow +}&\propto &
\sum_{{\bf R R}'}e^{i({\bf q}-{\bf q}_m)\cdot({\bf R}-{\bf R}')}\nonumber \\
&\sim&\sum_{\bbox \tau}\delta({\bf q}-{\bf q}_m-{\bbox \tau})	
\end{eqnarray}
for the neutron spin down to spin up channel. In these expressions
the nuclear reciprocal 
lattice vector ${\bbox \tau}$ and the spiral wave vector, ${\bf q}_m$,
are both parallel to the spiral axis.
Note that only one of the two magnetic
satellite Bragg peaks appears in each channel.
For a left-handed spiral, the signs of ${\bf q}_m$ in these expressions
reverse. Therefore  if the SDW in  $\rm V_{2-y}O_3$  is a spiral
with a macroscopic  handedness, we should find that 
magnetic Bragg  peaks with ${\bf Q}=(00\ell )$ 
come and go with a 180$^o$ rotation of the incident neutron spin state. 

\begin{figure}[bt]
\centerline{
\psfig{file=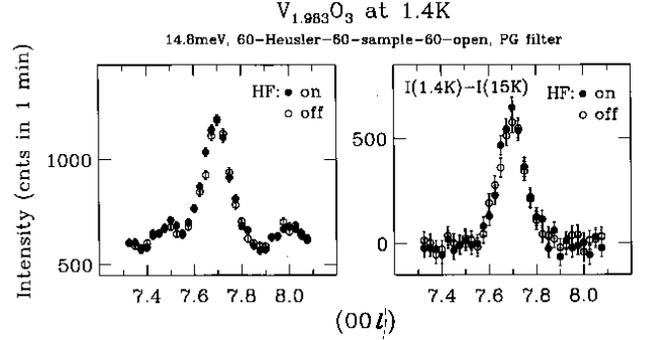,width=\columnwidth,angle=-90,clip=}}
\caption{(left) Double-axis polarized neutron scans at (0,0,7.7) 
with the neutron spin polarized along the {\em c} axis (HF) at 1.4K. 
(right) Identical scans at 15K were 
subtracted as a measure of  non-magnetic background.
The magnetic peaks are independent of the incident neutron spin state 
which shows that the SDW phase has no macroscopic handedness.
}
\label{polar2}
\end{figure}

Fig.~\ref{polar2} shows a double-axis polarized neutron scan through
the (0,0,7.7) magnetic Bragg peak of the SDW.  
The left panel shows raw data taken at 1.4~K, while in the right
panel, a non-magnetic background taken above the N\'{e}el temperature
has been subtracted.  Clearly the magnetic Bragg intensity does {\em
not} change upon flipping the incident neutron spin state. Hence we
conclude that the SDW phase does not have an intrinsic macroscopic
handedness. This means that if the ordered phase is a spiral, our
sample contains equal volume fractions of left and right handed
spirals. Another possibility is that we have an amplitude modulated SDW
which is a microscopic superposition of the left and right-handed
spirals.

\subsection{SDW, PM and AFI phase boundary at T=0}

\begin{figure}[bt]
\centerline{
\psfig{file=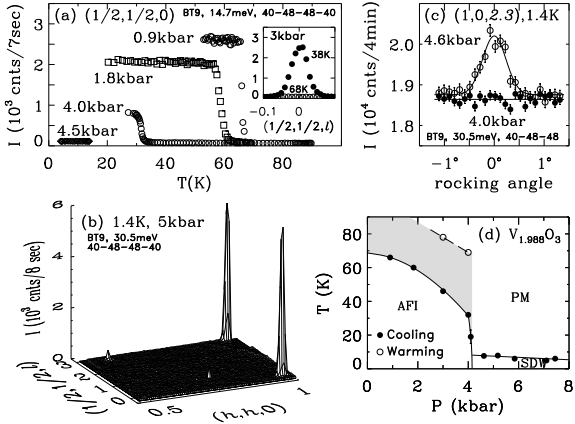,width=\columnwidth,angle=90,clip=}}
\caption{Phase transitions at high pressure for V$_{1.988}$O$_3$.
(a) Suppression of the AFI state by pressure, as measured by
monitoring the (1/2,1/2,0) magnetic Bragg peak upon cooling. 
Insert shows scans through (1/2,1/2,0)
at 3 kbar above (open circles) and below (solid circles) the AFI transition.
(b) Elastic scans at 1.4~K and 5.0 kbar, covering a region in the ($hhl$)
zone slightly larger than a quadrant of the Brillouin zone. 
Besides two spurious peaks which were found to be temperature
independent, there are only two resolution-limited 
nuclear Bragg peaks, (110) and (113). [No data were taken 
at the two corners near (1/2,1/2,3) and (113).]
(c) Elastic neutron scans at 1.4~K through (1,0,2.3) 
under pressures which stablizes the AFI (solid circles) and 
the SDW (open circles) ground states respectively.
(d) The pressure-temperature phase diagram for V$_{1.988}$O$_3$. 
The hatched region shows the thermal hysteresis at this first order transition.
}
\label{P_afi}
\end{figure}

After confirming the magnetic structure in the AFM phase
with polarized neutrons, let us now 
try to delimit the phase boundary of this SDW ground state in the
pressure-composition plane for V$_{2-y}$O$_3$. 
Two of the samples we studied with high pressure neutron diffraction
($y=0.012$ and 0.015)
have the AFI ground state at ambient pressure and display the metallic
state only under pressure. 
Fig.~\ref{P_afi} (a) shows the (1/2,1/2,0) Bragg peak which is the order parameter
of the AFI, as a function of temperature at various pressures
for a V$_{1.988}$O$_3$ sample.
Under high pressure, the transition to the AFI
remains strongly first-order,
as shown here by the sudden onset of antiferromagnetic
order with a nearly saturated magnetic moment.

The AFI state is completely suppressed above 4.1 kbar
for V$_{1.988}$O$_3$. A survey at 5 kbar
and 1.4~K in a quadrant of the ($hhl$) Brillouin zone
reveals no magnetic peaks associated with the AFI phase [refer to Fig.~\ref{P_afi} (b)].
On the other hand the Bragg peaks associated with SDW order are absent
for P=4.0 kbar but appear under a pressure of 4.6 kbar 
[refer to Fig.~\ref{P_afi} (c)].
The temperature-pressure phase diagram for V$_{1.988}$O$_3$
is shown in Fig.~\ref{P_afi} (d). The SDW transition is 
second order,\cite{bao93,bao96a} while the first order
AFI transition shows large thermal hysteresis.

\begin{figure}[bt]
\centerline{
\psfig{file=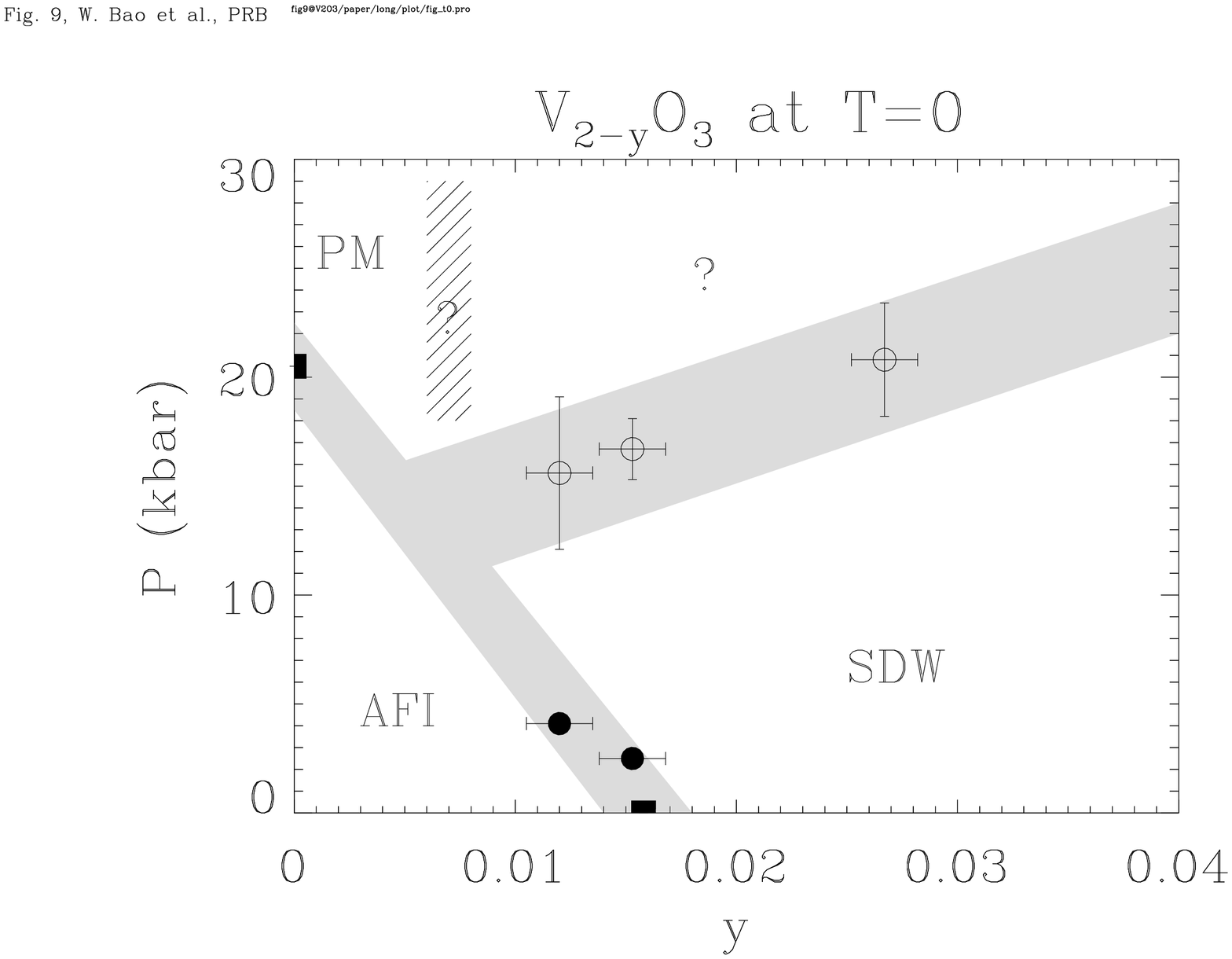,width=.7\columnwidth,angle=90,clip=}}
\caption{The composition-pressure phase diagram at $T=0$ for V$_{2-y}$O$_3$.
The squares are from [\protect\cite{bibdbma,bibsas}] and circles 
from the current neutron mesurement. 
Open circles are extrapolated from data below 8 kbar.
For $y\sim 0$, the paramagnetic metallic (PM) state is stable at
high pressures. However, the extent of the PM state beyond the hatched region
and down to temperatures below 0.1 K is
not clear.  It has not been possible to produce
single phase samples of $\rm V_{2-y}O_3$ for  $y>0.05$ because of the 
formation of V$_3$O$_5$ \protect\cite{bibyukk,bibsmplb}.
}
\label{grnd_state}
\end{figure}

Fig.~\ref{grnd_state} summarizes our current knowledge about ground
states in V$_{2-y}$O$_3$. For $y=0$, after the AFI is suppressed,
the metallic state is paramagnetic. For $y=0.012$ and 0.015, we found
that after the AFI is suppressed, the metallic state is a SDW. 
Up to 8 kbar,
the upper limit of our pressure cell, the SDW remains the stable 
low temperature state for all the samples ($y=0.012$, 0.015 and 0.027)
which we have studied with high pressure neutron diffraction\cite{bao93,bao96a}.
By {\em linear} extrapolation of the pressure dependent  N\'{e}el temperatures,
we derived estimates for the critical pressures of the SDW phase and these
are indicated by open circles in Fig.~\ref{grnd_state}. 

Anderson has pointed out the possibility
of a superconducting instability in the Brinkman-Rice liquid\cite{bibpwa}.
For metallic V$_{2-y}$O$_3$, SDW order appears to be 
the dominant instability.
However, it should be noted that the possibility of a superconducting 
state with $T_c<0.1$ K at high pressures has not been excluded yet\cite{newros}.

\section{Spin excitations and quantum critical behavior in metallic phases}

\subsection{Stoner electron-hole pair continuum at $T<T_N$}

\begin{figure}[bt]
\centerline{
\psfig{file=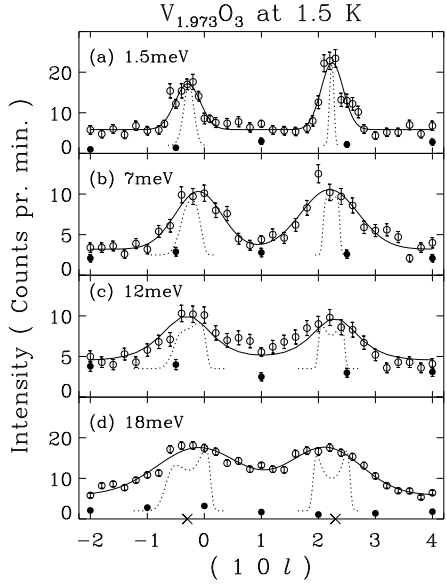,width=.8\columnwidth,angle=0,clip=}}
\caption{Constant energy, $\hbar\omega$, scans at 1.5~K for V$_{1.973}$O$_3$ 
across a  Brillouin zone. (10$\overline{2}$) and (104) are 
two nuclear Bragg points, and (1,0,$\overline{0.3}$) and (1,0,2.3) are 
two incommensurate magnetic satellites (marked by crosses). 
The spectrometer configurations
at BT2 were $E_f=13.7$ meV with horizontal collimations 60'-40'-40'-40'
for (a)-(c), and $E_i=35$ meV with 60'-40'-60'-60' for (d).
Vertical collimations were 300'-170'-230'-970' in
all four cases. The filled circles are analyzer turned backgrounds. The
dotted lines represent the four dimensional convolutions of the
instrumental resolution with an isotropic spin wave 
dispersion relation with a velocity $c=130$ meV$\AA$.  }
\label{afm_res}
\end{figure}

\begin{figure}[bt]
\centerline{
\psfig{file=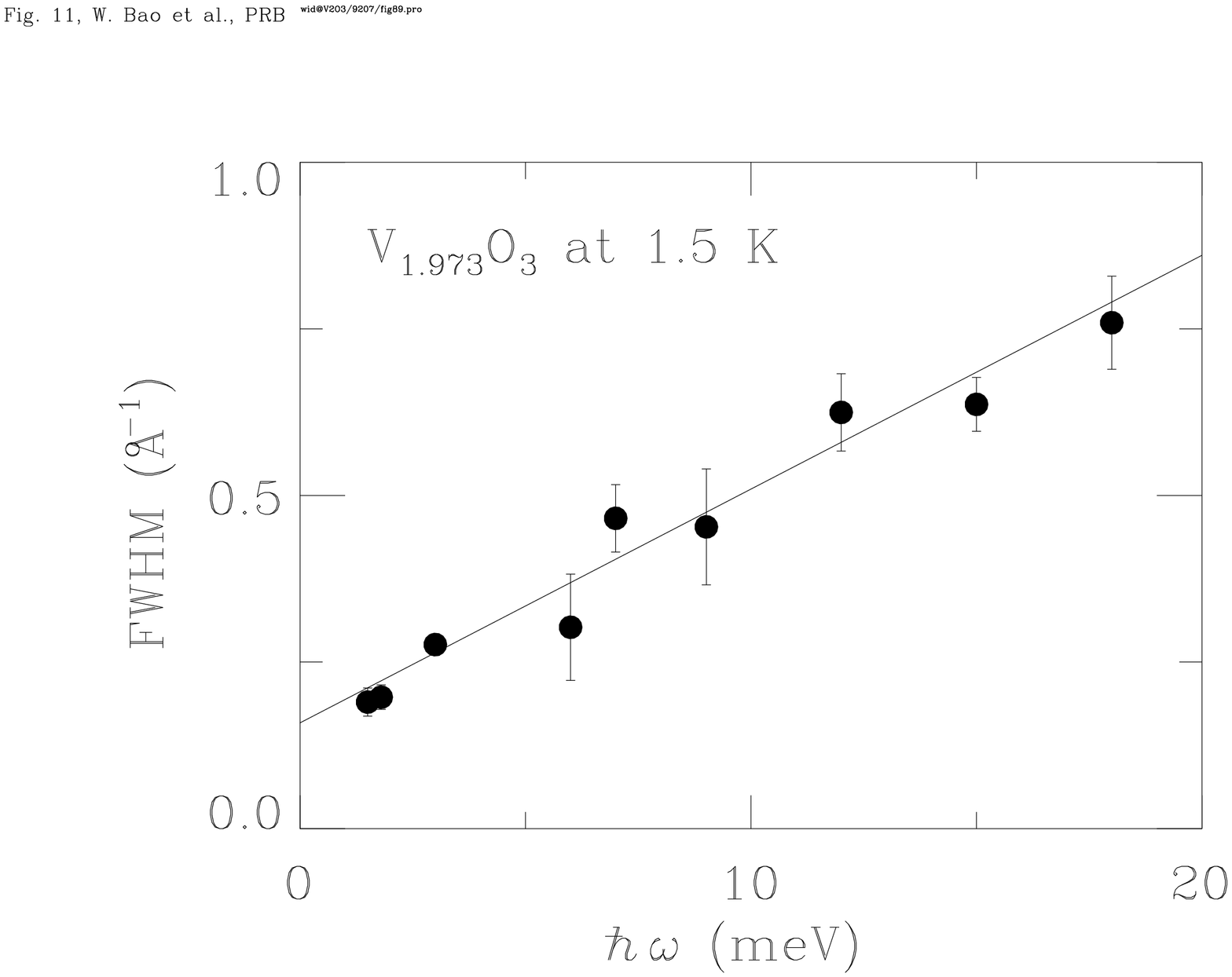,width=.8\columnwidth,angle=90,clip=}}
\caption{Resolution-corrected peak widths (FWHM) of constant energy scans
as a function of energy transfer for V$_{1.973}$O$_3$ in the long
range ordered SDW state at 1.5~K.}
\label{afm_wdth}
\end{figure}

Fig.~\ref{afm_res} shows constant energy transfer, $\hbar\omega$,
scans spanning a whole Brillouin zone in the 
AFM phase for V$_{1.973}$O$_3$ (T$_N=8.5(1)$~K). 
The incommensurate Bragg points, (1,0,$\overline{0.3}$) 
and (1,0,2.3), covered by these scans correspond to the
long range ordered SDW. The solid circles represent a 
flat background measured by turning the PG(002) 
analyzer 10$^o$ away from the reflection condition. 
Around each magnetic Bragg point, (1,0,$\overline{0.3}$) 
or (1,0,2.3),
there is only {\em one} broad peak. This is not due to poor
resolution merging two counter-propagating spin wave modes: 
the dotted lines in the figure represent the
calculated spectrometer response to spin waves  with a
velocity $c=130$ meV$\AA$, which is much larger than the inverse
slope of the peak width versus energy transfer curve 
(see Fig.~\ref{afm_wdth}). Our resolution is clearly adequate to rule
out a conventional spin wave response in this energy range. Therefore the
broad intensity profile which we observed 
is intrinsic, and in the energy range probed,
magnetic excitations consist of a {\em single} broad lobe at each
magnetic Bragg point.

For insulating magnets, propagating spin waves exist over the whole
Brillouin zone. For itinerant magnets, however,  propagating spin waves
only exist in a limited range near the magnetic Bragg peaks.
Beyond this range they merge into the Stoner particle-hole
continuum. According to weak coupling theory,\cite{sdw_paf} 
the energy gap which a SDW creates inside the particle-hole 
continuum is $\Delta =1.76\alpha k_BT_N$, where
$\alpha \equiv (v_1+v_2)/2(v_1v_2)^{1/2}$ and $v_1$ and $v_2$
are Fermi velocities of the two nesting bands. 
Inside this gap, spin waves propagate with
a velocity\cite{sdw_paf,Cr_shl,Cr_jbs}
$c=(v_1v_2/3)^{1/2}$.
For metallic V$_{2-y}$O$_3$ at T=0, spin waves are expected only below
$\sim$$1.76k_B T_N = 1.4$~meV, which is below the energies which
we have probed in this experiment. 
The broad lobes measured in Fig.~\ref{afm_res} thus represent
magnetic excitations inside the Stoner continuum.

Other signs that we are probing magnetic fluctuations in the electron-hole
continuum, as opposed to transverse oscillations
of the order parameter, are (1)
The spectral weight from 1.5 meV to 18 meV in Fig.~\ref{afm_res}
corresponds to a fluctuating moment of
0.32$\mu_B$ per V, already twice that of the 
static staggered moment,\cite{bao93} and
(2) The resolution corrected full-widths-at-half-maximum (FWHM) of constant 
energy scans (see Fig.~\ref{afm_wdth}) 
extrapolate to a finite value at $\omega=0$, even though there are
true, resolution-limited, Bragg peaks at $\omega=0$.

For Cr metal\cite{Cr_ordr,Cr_rev}, another 3-D SDW (T$_N=310$~K),
one would expect at T=0 that spin wave modes merge into
a single broad lobe above $\sim$1.76$k_BT_N$=47 meV. 
However, due to the much larger spin wave velocity in Cr, 
estimated at $\sim$1000 meV$\AA$\cite{Cr_shn,Cr_aas},
it has not yet been possible in this material to 
distinguish between
resolution merged spin waves and single lobe excitations. 
We were aided in the case of
metallic V$_2$O$_3$ by strong correlations. The enhanced effective
mass reduces the Fermi velocity to a value which is $\sim 1/15$
of that for Cr.\cite{bao93} This makes it feasible
to unambiguously observe
these unique single-lobe magnetic excitations in the SDW state of
V$_{2-y}$O$_3$.

\subsection{Doping effects in the metallic phases}

\begin{figure}[bt]
\centerline{
\psfig{file=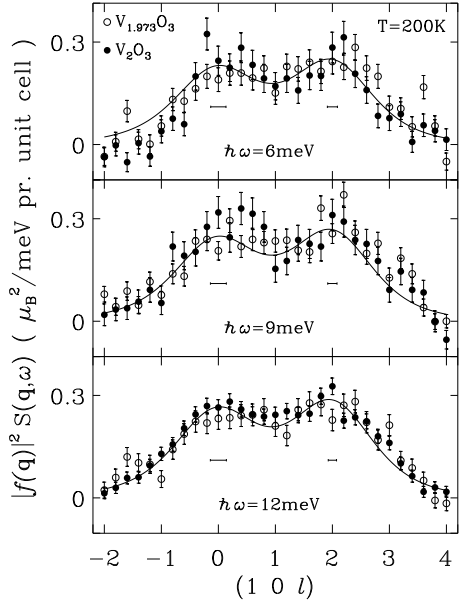,width=.8\columnwidth,angle=0,clip=}}
\caption{Magnetic responses at 200~K from V$_{1.973}$O$_3$ (open circles)
and V$_2$O$_3$ (solid circles) in the PM phase. Although these two samples
have different low temperature states, the SDW for the former and the 
AFI for the
latter, there is no difference in the dynamic correlations in the PM phase.
The experimental configuration was E$_f=13.7$ meV with horizontal collimations
60'-40'-40'-60' for V$_{1.973}$O$_3$ at BT2. 
The horizontal bars indicate the FWHM of the projection
of the resolution function on the scan direction.
The experimental configuration for V$_{2}$O$_3$ were E$_f=13.7$ meV 
with horizontal collimations 40'-60'-80'-80' for $\hbar\omega=6$ and 9~meV,
and E$_i=35$meV with 40'-40'-80'-80' for 12~meV at BT4 with PG monochromator
and analyzer. The solid line is a 
least-squares fit to Eq.\ (\ref{af_scr1})
and (\ref{af_scr2}).}
\label{pure_doped}
\end{figure}

We have previously shown\cite{bao93} that both the staggered moment
and the incommensurate wave vector of the static SDW order depend only weakly
on doping. For a change in metal-deficiency 
$\delta y=0.03$ which is twice the critical $y_c$ for the AFI phase,  
the change in the magnetic wave vector (1.7{\bf c}$^*$) 
is only $\sim$0.02{\bf c}$^*$.
Likewise, dynamic magnetic correlations in the PM phase 
are insensitive to doping. As shown in Fig.~\ref{pure_doped},
even for pure V$_2$O$_3$ which has different low temperature
AFI state, the magnetic response at 200 K is indistinguishable from
that of V$_{1.973}$O$_3$. In particular we do observe peaks 
at the SDW wave vectors.
This demonstrates that magnetism throughout the metallic phase
is controlled by the same Fermi surface instability and indicates that 
small moment incommensurate 
antiferromagnetism in $\rm V_{2-y}O_3$ is not due to impurity effects but 
is an 
intrinsic property of metallic $\rm V_2O_3$.

\subsection{Antiferromagnetic quantum critical behavior in metal} 

The SDW in metallic V$_{2-y}$O$_3$ is a small moment, low temperature
antiferromagnetic ordering. The staggered moment 
in the magnetic ground state (0.15$\mu_B$) is only a
small fraction of the effective paramagnetic moment (2.37-2.69$\mu_B$) 
which appears in the high temperature Curie-Weiss 
susceptibility\cite{bibedjx,bibdbme,bibyukkd}.
In other words, the magnetic system
is very close to a quantum critical 
point\cite{ja_hertz,2dheis,2dheiqc,isfandy}.
We have previously shown\cite{bao96a} that the single lobe 
spin excitation spectrum in
metallic V$_{1.973}$O$_3$ is isotropic in space, and the spectrum 
$\chi''({\bf q},\omega)$, from
1.4K inside the ordered phase to temperatures above 20T$_N$, 
has been determined in absolute units\cite{e_used}. Here we discuss our
results in connection with the quantum critical behavior.
 
\begin{figure}[bt]
\centerline{
\psfig{file=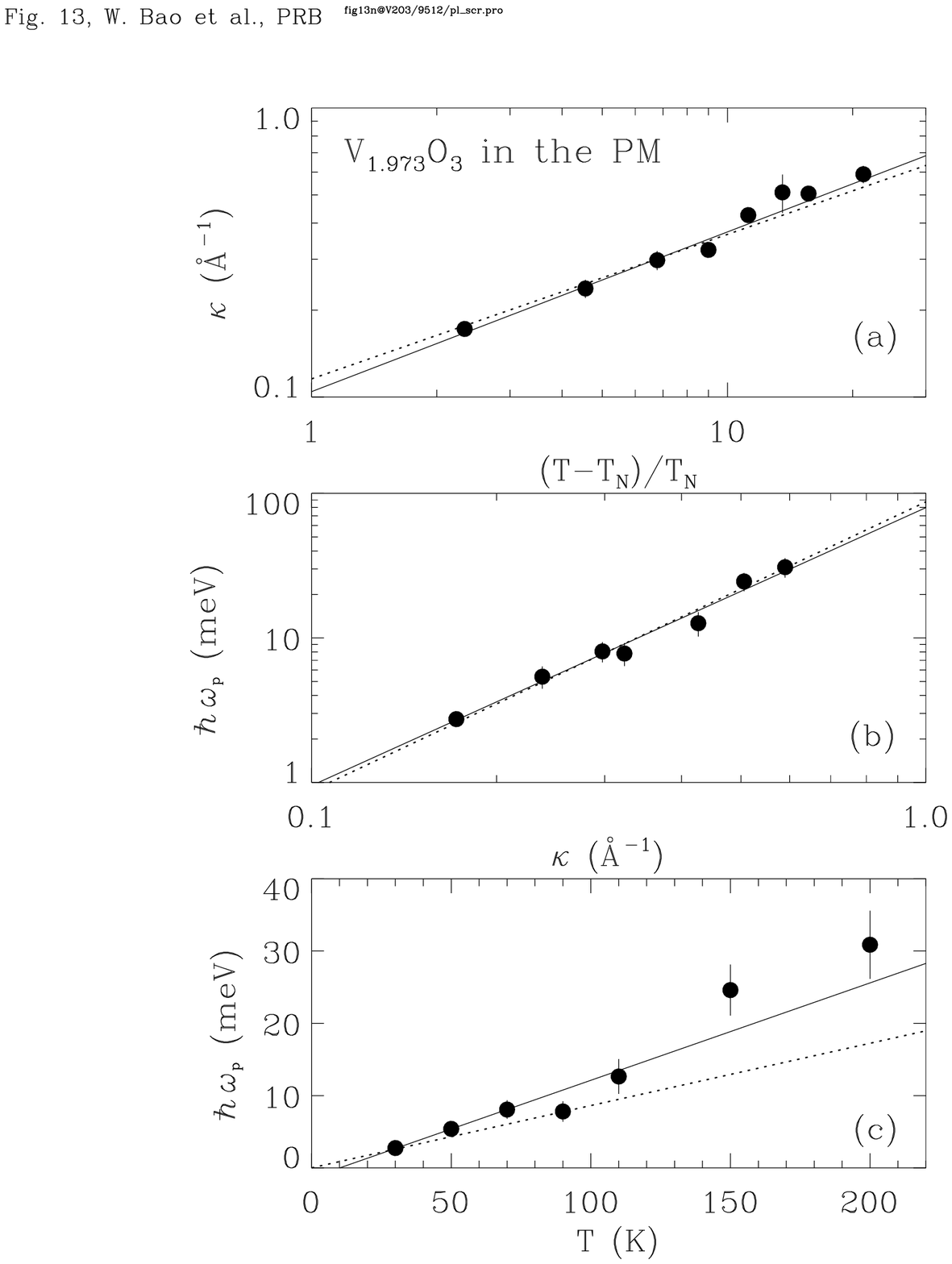,width=.8\columnwidth,angle=0,clip=}}
\caption{(a) The inverse correlation length as a function of the
reduced temperature for V$_{1.973}$O$_3$ in the PM phase.
Both axes use the logarithmic scale. The solid line is
the best fit of $\kappa \propto t^\nu$ to the data with
$\nu=0.55(4)$, and the dotted line is a fit with fixed $\nu=1/2$.
(b) The characteristic spin fluctuation energy, $\hbar\omega_p$, where
$\chi''({\bf Q},\omega)$ is maximized, as a function of $\kappa$.
The solid line is
the best fit of $\omega_p \propto \kappa^z$ to the data with
$z=1.9(1)$, and the dotted line is a fit with fixed $z=2$.
(c) $\hbar\omega_p$ is roughly proportional to the thermal 
energy $k_B T$ (the dotted line). The solid line is 
$\omega_p \propto T-T_N$.
}
\label{gk2}
\end{figure}

Fig.~\ref{gk2}(a) shows in a log-log scale
the inverse correlation length, $\kappa$,
measured in Ref.\ [\cite{bao96a}] 
as a function of the reduced temperature, $t\equiv (T-T_N)/T_N$.
The solid line is $\kappa\propto t^{\nu}$
with $\nu=0.55(4)$ from a least squares fit. This value of $\nu$
suggests a small positive $\eta$ exponent according to 
the scaling relation\cite{fisher} $\eta=2-\gamma/\nu$. 
However, the classical values
of $\nu=1/2, \eta=0$ are also consistent with our data 
(refer to the dotted line).

The spin fluctuation spectrum $\chi''({\bf q},\omega)$
reaches its maximum value, $\chi''_{max}$,
at $\omega=\omega_p$ and ${\bf q=Q}$ which is an antiferromagnetic
Bragg point\cite{bao96a}.
The dominant part of the spectral weight, therefore, falls
in the hydrodynamic region\cite{crt_hh}, $|{\bf q-Q}|\ll \kappa$,
where $\hbar \omega_p =\gamma_{A}\, \kappa^z$. 
In Fig.~\ref{gk2}(b),
we find the dynamic exponent $z=1.9(1)$ (the solid line), 
which is in agreement with the theoretical prediction $z=2$ 
for itinerant antiferromagnets in the
quantum critical region\cite{ja_hertz,isfandy}. The constant prefactor is 
$\gamma_A=87(4)$ meV$\AA^2$.

Combining the scaling relations from Fig.~\ref{gk2}(a) and (b),
we have $\hbar\omega_p \sim  t^{z\nu} = C k_B (T-T_N)$
[refer to the solid line in Fig.~\ref{gk2}(c)].
Given the small value of $T_N$ and the fact that $C\approx 1$, 
the characteristic spin fluctuation energy $\hbar\omega_p$
is close to $k_B T$ [refer to the dotted line 
in Fig.~\ref{gk2}(c)]. This feature is shared with nearly
metallic cuprates\cite{la2smha} and some heavy fermion systems\cite{qcp_as}.
Antiferromagnetic spin fluctuations for systems
close to a quantum critical point, therefore, might provide a
microscopic mechanism for marginal Fermi liquid 
phenomenology\cite{bibmfl}.

\begin{figure}[bt]
\centerline{
\psfig{file=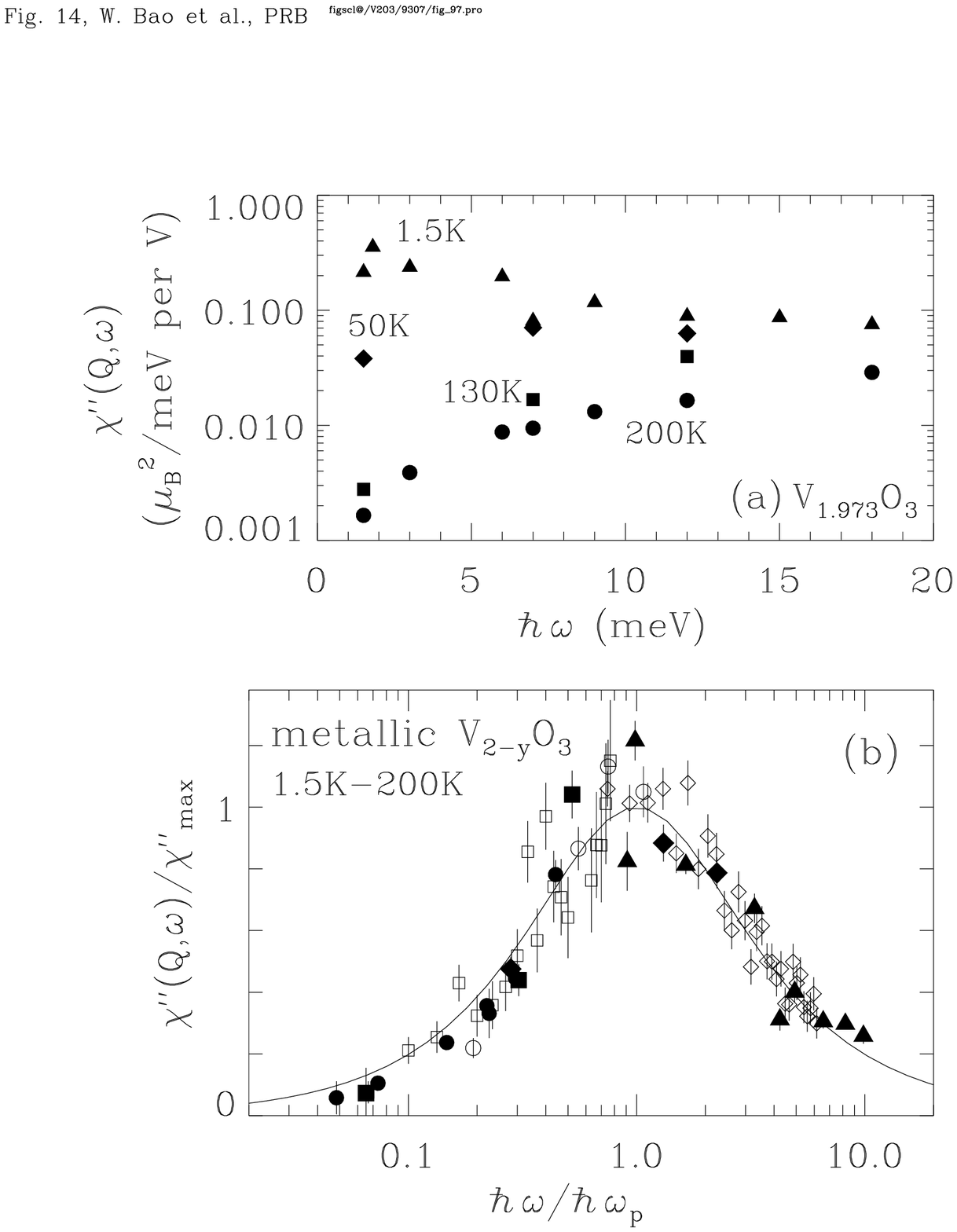,width=.9\columnwidth,angle=0,clip=}}
\caption{(a) Normalized peak intensity of constant energy scans
at 1.5K (triangles), 50K (diamonds), 130K (squares) and 200K (circles)
for V$_{1.973}$O$_3$. In the measured
energy range, $\chi''({\bf Q},\omega)$ changes from a decreasing function
of $\hbar \omega$ at 1.5K to an increasing function at 200K. (b) Data
in (a) are scaled onto a universal curve using measured\protect\cite{bao96a}
$\chi''_{max}=\chi_{\bf Q}/2$ and $\hbar\omega_p=\gamma_A\kappa^2$. 
The open circles are from constant-$\hbar \omega$=1.5meV 
scans at 9K, 19K, 27K and 82~K; the open diamonds are from a
constant-{\bf Q}=(1,0,2.3) scan at 30~K. 
Constant-{\bf Q} scan for V$_2$O$_3$
at 200~K is shown with the open squares.
The solid line
is $z=2y/(y^2+1)$ [refer to Eq.\ (\ref{eq_afcont})
and (\ref{eq_sclaf})].
}
\label{scr_scl}
\end{figure}

The peak intensity of some constant energy scans is shown in 
Fig.~\ref{scr_scl}(a) as function of energy. 
They are expressed in absolute units and 
represent $\chi''({\bf Q},\omega)$ at the magnetic zone center. 
At 1.5K, $\chi''({\bf Q},\omega)$ is a decreasing function
of $\omega$ in the measured energy range and it changes to an increasing
function at elevated temperatures. They were scaled onto a single universal
curve in Fig.~\ref{scr_scl}(b) by dividing $\chi''({\bf Q},\omega)$ 
by $\chi''_{max}$ and $\hbar\omega$ by the peak energy 
$\hbar\omega_p$. 
The apparently different behaviors in a {\em finite} energy window in 
Fig.~\ref{scr_scl}(a), therefore, reflect different segments of 
a universal behavior.

For a general wave vector, ${\bf q}$, away from an antiferromagnetic
Bragg point, ${\bf Q}$, we have shown\cite{bao96a} 
previously
that our neutron scattering data measured at various temperatures
can be described by
\begin{equation}
\frac{\chi''({\bf Q+q},\omega)}{2 \chi''_{max}}=
\Phi\left(\frac{q}{\kappa}, \frac{\hbar\omega}{\gamma_A\, \kappa^2}\right),
\label{eq_afcont}
\end{equation}
where the scaling function is 
\begin{equation}
\Phi(x,y)=y/[y^2+(1+x^2)^2]. \label{eq_sclaf} 
\end{equation}
The solid line in Fig.~\ref{scr_scl}(b) is $2\Phi(0,y)$.
The maximum $\chi''_{max}$ is related to
static staggered susceptibility\cite{bao96a}.
Combining Kramers-Kronig relation and Eq.\ (\ref{eq_afcont}) gives
\begin{eqnarray}
\chi_{\bf Q} &\equiv & \pi^{-1} \int_{-\infty}^{\infty} d\omega 
\chi''({\bf Q},\omega)/\omega \nonumber \\
&=& \chi''_{max} 2\pi^{-1}  \int_{-\infty}^{\infty} 
\Phi(0,\hbar\omega/\gamma_A \kappa^2)\,
d\omega/\omega \nonumber \\
&=& \chi''_{max} 2\pi^{-1} \int_{-\infty}^{\infty} \Phi(0,y)\, 
dy/y \nonumber \\
&=&\chi''_{max} \times constant.  
\end{eqnarray}
Notice that when $T_N\rightarrow 0$,
$\hbar\omega_p = Ck_B T$,
$\chi''_{max}\sim \chi_{\bf Q} = a_1 T^{-\gamma}$ 
and $\kappa= T^{\nu}/a_2$, 
hence Eq.\ (\ref{eq_afcont}) becomes
\begin{equation}
\chi''({\bf Q+q},\omega)=\frac{a_1}{T^{\gamma}}\,
\Phi\left(\frac{a_2 q}{T^{\nu}}, \frac{\hbar\omega}{C k_B T}\right),
\label{eq_junk}
\end{equation}
where $a_1$ and $a_2$ are non-universal constants.\cite{notea}
Using $z\nu\approx 1$ from our measurement and the scaling relation
$\gamma=(2-\eta)\nu$, and absorbing the constant $C$ in the 
scaling function,
\begin{equation}
\chi''({\bf Q+q},\omega)=\frac{a_1}{T^{(2-\eta)/z}}
\Phi_1\left(\frac{a_2 q}{T^{1/z}}, \frac{\hbar\omega}{k_B T}\right).
\label{eq_sy}
\end{equation}
This expression, which is supported by our
experimental data, is identical to the scaling form proposed 
by Sachdev and Ye in the quantum critical region\cite{2dheiqc}.

\subsection{Theory of itinerant magnetism} 

Itinerant magnetic systems have been studied
within the random-phase approximation (RPA), which provides a
reasonable description at T=0 for the ordered moment and for
spin excitations in the spin wave modes and in the Stoner continuum.
But it does not relate the low-temperature excitation parameters
to finite-temperature properties such as
the ordering temperature and the high temperature Curie-Weiss 
susceptibility. 

The imaginary part of the RPA generalized
susceptibility for an itinerant antiferromagnet, 
as was done to describe magnetic fluctuations in 
chromium\cite{Cr_shl,Cr_jbs}, can be written as follows:
\begin{equation}
\chi''({\bf Q+q},\omega)=
	\frac{\chi_0/(\kappa \xi_0)^2}{1+({\rm q}/\kappa)^2}\,
	\frac{\hbar\omega\Gamma({\rm q},\kappa)}{(\hbar\omega)^2
	+\Gamma({\rm q},\kappa)^2}, \label{af_rpa} 
\end{equation}
where the wave vector dependent relaxation rate,
\begin{equation}
\Gamma({\rm q},\kappa)=\gamma_A\,(\kappa^2+{\rm q}^2),
\label{af_scr2}
\end{equation}
corresponds to the simplest
functional form consistent with the dynamical scaling hypothesis
with a dynamic exponent $z=2$.  As a comparison, 
for the insulating Heisenberg 
antiferromagnet,\cite{crt_hh,crt_lau} $z=3/2$.
The uniform non-interacting susceptibility $\chi_0$,
the length scale $\xi_0$, and $\gamma_A$ are 
temperature-independent {\em parameters}, and the inverse correlation
length $\kappa$ approaches zero when the N\'{e}el temperature
is approached. This functional form and derivations 
thereupon\cite{Cr_sato}
offer a good description of critical
spin fluctuations for Cr and its alloys\cite{Cr_aas,Cr_nhf}.
They have also been successfully used to describe spin dynamics 
as measured by neutron scattering for La$_{1.86}$Sr$_{0.14}$CuO$_2$
above the superconducting transition temperature\cite{la2tem}.
A modified form was used to model nuclear magnetic resonance (NMR)
data in other high temperature superconducting 
cuprates\cite{rpa_mmp,rpa_bhsb,rpa_mpt}. However, the
RPA parameters $\gamma_0$, $\chi_0$,
and $\xi_0$ in Eq.\ (\ref{af_rpa}) are free parameters which
are not connected to other physically measurable quantities. 

The SCR theory,\cite{scrmd,scrmor0,scrrama,scrlnzr,scrmor} 
a mode-mode coupling theory,
improves upon the RPA theory by considering  
self-consistently the effects of spin fluctuations on the thermal 
equilibrium state. It is solvable in the small moment limit with
a few experimentally measurable parameters. 
The ferromagnetic version of the SCR theory was discussed 
in [\cite{scrmd,scrmor0,scrrama,scrlnzr}] 
and the antiferromagnetic version 
in [\cite{scrmor,scrmorc}]. 

The generalized susceptibility for antiferromagnets
in the SCR theory\cite{scrmorc} has the same functional
form as in the RPA theory:
\begin{equation}
\chi''({\bf Q+q},\omega)=
	\frac{\chi_{\bf Q}}{1+({\rm q}/\kappa)^2}\,
	\frac{\hbar\omega\Gamma({\rm q},\kappa)}{(\hbar\omega)^2
	+\Gamma({\rm q},\kappa)^2}, \label{af_scr1} 
\end{equation}
where $\Gamma({\rm q},\kappa)$ is defined as in Eq.\ (\ref{af_scr2}).
The RPA parameter $\chi_0/(\kappa \xi_0)^2$
is substituted now by the static staggered susceptibility $\chi_{\bf Q}$,
in consistence with the Kramers-Kronig relation.
Major difference from the RPA theory, however, is that
the SCR theory can predict 
other physical quantities using
parameters from spin excitation spectrum\cite{scr_hh}. For example, 
the non-universal parameters\cite{notea} $a_1$ and $a_2$ defined above 
Eq. (\ref{eq_junk}) can be used to {\em quantitatively} 
relate $T_N$ to the staggered moment $M_{\bf Q}$ 
(refer to Fig.\ 4 of Ref.\ [\cite{bao96a}]).

\begin{figure}[bt]
\centerline{
\psfig{file=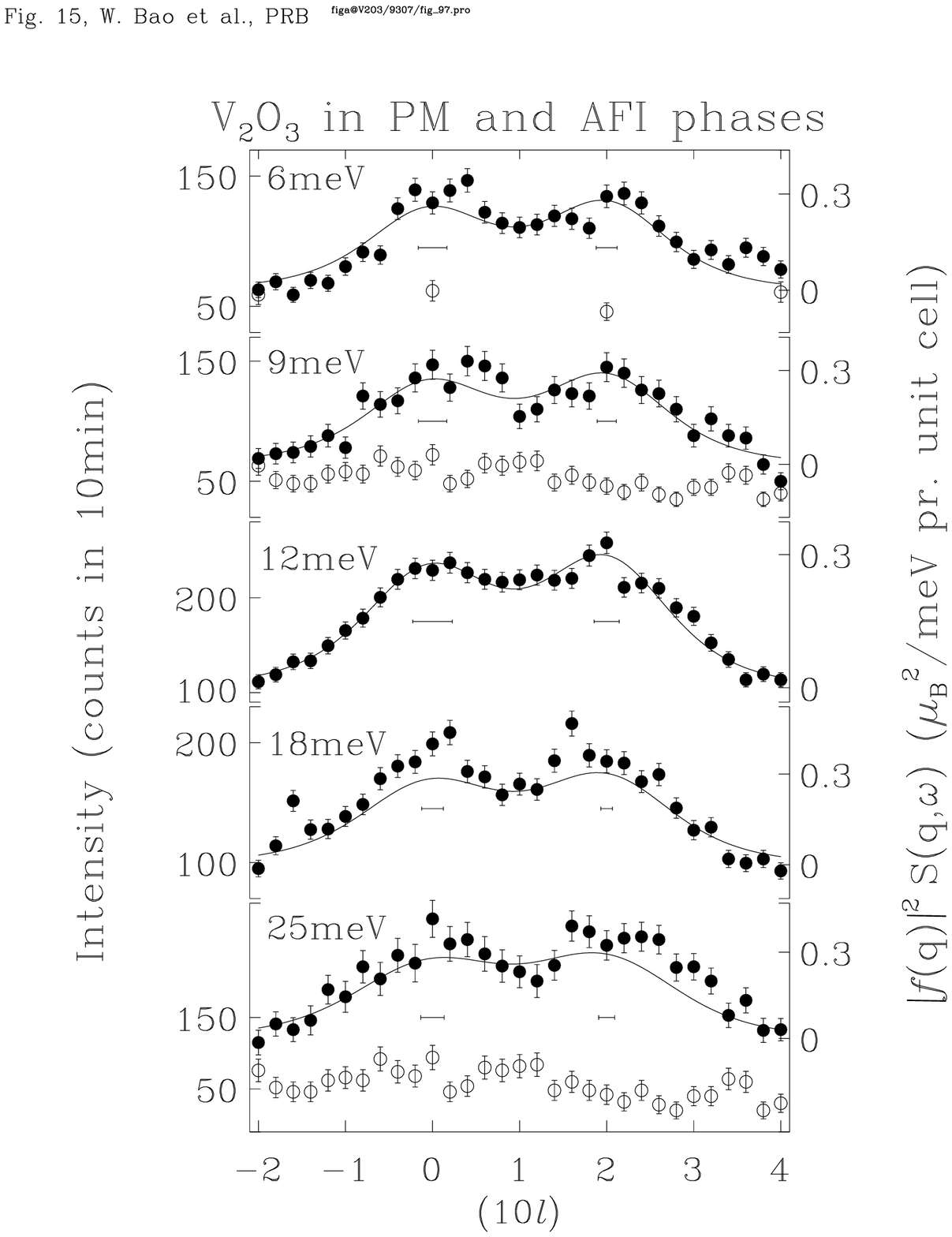,width=.7\columnwidth,angle=0,clip=}}
\caption{Constant energy scans at 200~K in the PM phase (solid circles)
and at 160~K in the AFI phase (open circles) for stoichiometric V$_2$O$_3$.
Magnetic correlations near the SDW wave vectors, represented by the
two peaks, disappear when the material enters the AFI phase. The experimental
configurations at BT4 were E$_f=13.7$ meV with horizontal collimations 
40'-60'-80'-80' for $\hbar\omega=6$ and 9~meV,
E$_i=35$meV with 40'-40'-80'-80' for 12 and 18~meV, and
E$_f=30$meV with 60'-60'-60'-60' for 25~meV. We used PG (002) 
as monochromator
and analyzer except for the last configuration where
a Cu (220) monochromator was used to improve the energy resolution. 
The horizontal bars indicate the FWHM of 
the projection
of the resolution function on the scan direction.
The solid line is a least-squares fit to Eq.\ (\ref{af_scr1})
and (\ref{af_scr2}) and
absolute units are given on the right scales.}
\label{fig_paf}
\end{figure}

Fig.~\ref{fig_paf} shows an example of a simultaneous fit of
constant energy scans to Eq.\ (\ref{af_scr1}) and (\ref{af_scr2})
for stoichiometric V$_2$O$_3$ at 200K. 
Solid circles represent the magnetic response in
the paramagnetic metallic phase, which peaks around the SDW wave vectors.
The resolution widths are much smaller than the measured widths,
therefore  resolution broadening can be ignored in this case.
The solid line through these 200K data is the
best-fit functional form for $|f({\bf q})|^2 {\cal S}({\bf q},\omega)$ 
calculated using copies of Eq.\ (\ref{af_scr1}) summed over the 
Bragg points ${\bf Q}=(1,0,4-\Delta)$ 
and $(1,0,\overline{2}+\Delta)$ with
$\Delta=1.95(6) c^*$, $\kappa=
1.32(5) c^*$, $\gamma_{A}=18(2)$ meV/$c^{*2}$ and 
$\chi_{\bf Q}=0.071(7) \mu_B^2$/meV per vanadium.
These values are within the error-bars of those for 
V$_{1.973}$O$_3$,\cite{bao96a}
as should be expected given the close similarity of the raw data
for these two samples which are shown in 
Fig.~\ref{pure_doped}.

\begin{figure}[bt]
\centerline{
\psfig{file=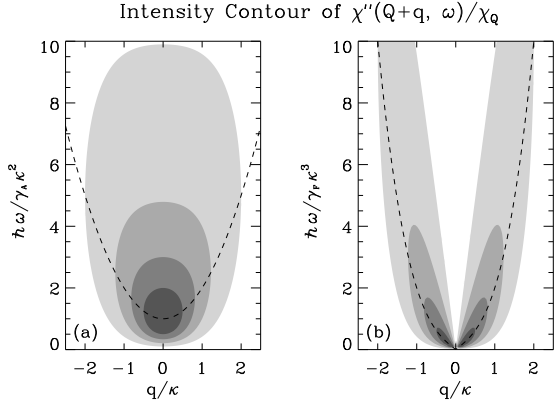,width=\columnwidth,angle=90,clip=}}
\caption{Intensity contour of the scaled magnetic response function 
$\chi''({\bf Q+q},\omega)/\chi_{\bf Q}$ as function of reduced wave 
vector and reduced energy for
(a) itinerant antiferromagnets and for
(b) itinerant ferromagnets near a quantum critical point.
The contour levels are 0.1, 0.2, 0.3
and 0.4. In the antiferromagnetic case (a), the peak intensity
is 0.5 at q=0 and $\hbar \omega/\gamma_{A} \kappa^2$=1. In the ferromagnetic
case (b),
the intensity is singular at q=0 and $\omega$=0. Both $\kappa$ and 
$\chi_{\bf Q}$ are functions of temperature. $\gamma_{A}$ 
and $\gamma_{F}$ are temperature
insensitive constants. See text for details.
}
\label{scr_cnt}
\end{figure}

It is customary to use the Lorentzian relaxation energy
$\Gamma({\rm q},\kappa)$ of (\ref{af_rpa}) or (\ref{af_scr1})
as the energy scale of spin fluctuations. 
For an itinerant antiferromagnet
close to a quantum critical point, such as metallic V$_{2-y}$O$_3$,
this reduces to $\gamma_A \kappa^2$ at $q=0$\cite{bao96a}.
To demonstrate the prominence of this energy, 
we plot in Fig.~\ref{scr_cnt}(a)
the intensity contour of $\chi''({\bf Q+q},\omega)/\chi_{\bf Q}
=\Phi(q/\kappa, \hbar\omega/\gamma_A\, \kappa^2)$ 
for small moment itinerant antiferromagnets 
[refer to Eq. (\ref{af_scr1}), (\ref{af_scr2}) and (\ref{eq_sclaf})].
The spectrum is unique in that the electron-hole damping has 
reduced the excitation spectrum
to a single lobe at the magnetic zone center, as we observed in
our neutron scattering experiment for metallic 
V$_{2-y}$O$_3$\cite{bao93,bao96c}. The peak intensity occurs at
$q=0$ and $\hbar\omega = \gamma_A\kappa^2$, 
and $\gamma_A\kappa^2\sim t^{z\nu}$
clearly is the characteristic energy scale.
In this case, $z\nu\approx 1$, as we found experimentally, is
essential for achieving the quantum critical scaling of (\ref{eq_sy}).

This is quite different from the well studied case of small moment
itinerant ferromagnets.
The $\chi''({\bf q},\omega)$ predicted by the SCR 
theory for ferromagnets\cite{NA_ggl,scrmnsid} has the 
same Lorentzian form as (\ref{af_scr1}),
but the relaxation energy is given by
\begin{equation}
\Gamma({\rm q},\kappa)=\gamma_F\,{\rm q}(\kappa^2+{\rm q}^2). \label{fm_scr}
\end{equation}
Like the antiferromagnetic SCR theory, the ferromagnetic SCR theory
not only correctly describes the spin fluctuation spectrum, but also
quantitatively relates the spectrum parameters to various other 
physical quantities, as demonstrated in the 3-D small-moment 
ferromagnets MnSi\cite{MnSi_ishc,scrmnsid,MnSi_ishii}
and Ni$_3$Al\cite{NA_ggl}.
Recasting Eq.\ (\ref{af_scr1}) and (\ref{fm_scr}) in a dimensionless form,
\begin{equation}
\chi''({\bf Q+q},\omega)/\chi_{\bf Q}=\Psi\left(\frac{q}{\kappa}, 
\frac{\hbar\omega}{\gamma_F\, \kappa^3}\right),
\end{equation}
with
\begin{equation}
\Psi(x,y)=|x|y/[y^2+(x+x^3)^2]. \label{eq_fmcont}
\end{equation}
It is depicted in Fig.~\ref{scr_cnt}(b).
Although the counter-propagating 
spin wave branches are over-damped by electron-hole 
excitations, they remain two separate branches. The intensity
is singular at $q=0$ and $\omega=0$, and the Lorentzian
relaxation energy (the dashed line) serves well as the energy scale
in this case, as customarily performed.

\section{Spin waves in the AFI phase}
\label{sec_afi}

While magnetic excitations in the metallic phase are dominated
by over-damped modes,
the AFI phase exhibits conventional propagating spin wave excitations.
Word et al.\ \cite{bibrew} measured spin waves in this phase
around (1/2,1/2,0) at 142K in pure V$_2$O$_3$.
There is a spin excitation gap of 4.8 meV, which is essentially
independent of temperature for $T<T_N$\cite{bibmysawb}. 
For V$_{1.94}$Cr$_{0.06}$O$_3$ (T$_N$=177K), the gap energy
at (1/2,1/2,0) is not different from that seen for
the pure sample for temperatures
below 50K\cite{bibmysawb}.
However, it shows substantial softening with rising temperature,
consistent with the trend towards a continuous phase transition with
increasing Cr doping.

\begin{figure}[bt]
\centerline{
\psfig{file=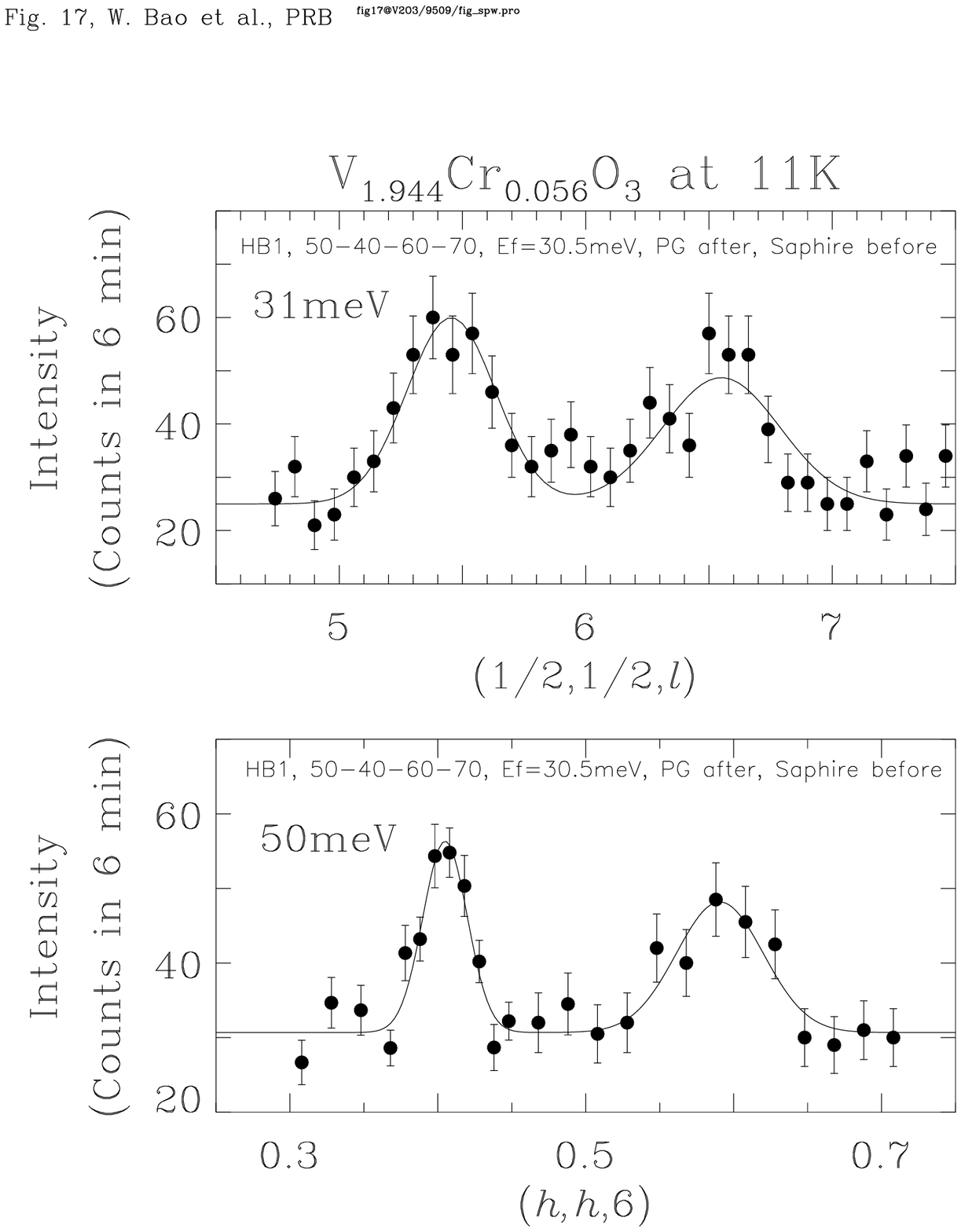,width=.7\columnwidth,angle=0,clip=}}
\caption{Constant energy transfer scans along the (001) and (110) directions
around the magnetic Bragg point (1/2,1/2,0). The two peaks in each frame
correspond to counter-propagating spin wave modes. 
The different peak shapes are due to resolution effects 
(focusing and defocusing).
}
\label{afi_scan}
\end{figure}

\begin{figure}[bt]
\centerline{
\psfig{file=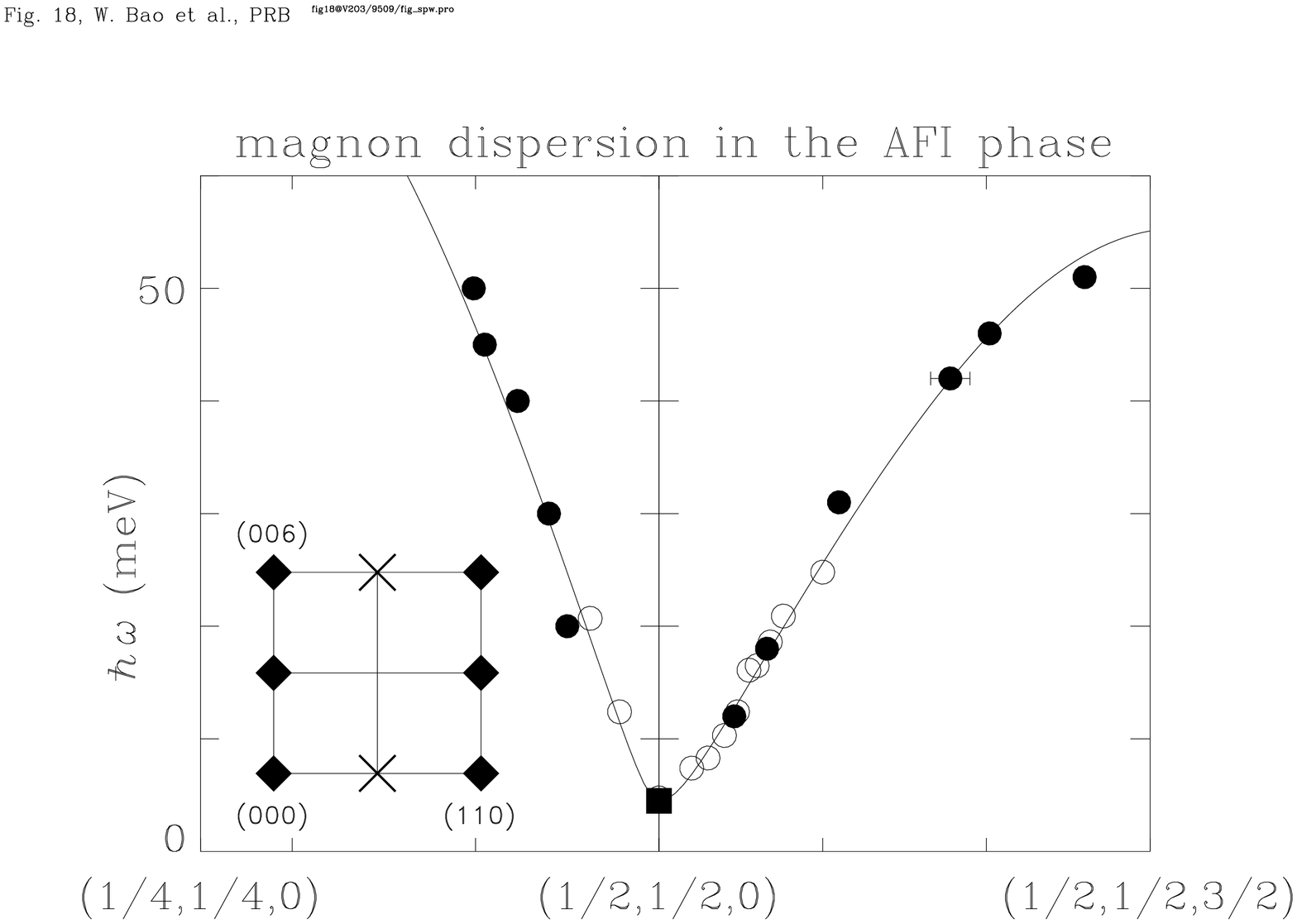,width=.9\columnwidth,angle=90,clip=}}
\caption{Spin wave dispersion near the antiferromagnetic zone center.
Solid circles represent the current measurements at 11~K for 
V$_{1.944}$Cr$_{0.056}$O$_3$, the square is from Ref. 
[\protect\cite{bibmysawb}]. For comparison, data from 
[\protect\cite{bibrew}] for pure V$_2$O$_3$ at 142~K are
represented by open circles.
With these limited data, individual exchange constants cannot be 
derived reliably. Only a specific linear combination as described 
in the text can be determined. Insert: the reciprocal lattice
plane for the measurement. The diamonds denote nuclear Bragg points,
and crosses magnetic Bragg points.
}
\label{fig_disp}
\end{figure}

Here we present spin waves measured at 11~K for 
V$_{1.944}$Cr$_{0.056}$O$_3$.
Fig.~\ref{afi_scan} shows constant energy scans
near a magnetic zone center. The spin wave dispersion
relation has 
been measured for energies up to 50 meV (see Fig.~\ref{fig_disp})
which covers about one quarter of the Brillouin zone.
For comparison, some of previous data
for pure V$_2$O$_3$ are also included in Fig.~\ref{fig_disp}.
At low measurement temperatures and for $\hbar\omega <25$ meV, 
there is no difference in the spin wave dispersion for pure and 3\% Cr doped
V$_2$O$_3$.

Spin waves in the isomorphous corundum
antiferromagnets Cr$_2$O$_3$ and $\alpha$-Fe$_2$O$_3$
have been measured over the whole 
Brillouin zone\cite{bibejsa,bibejsb}. These two compounds
have different magnetic structures\cite{bibbnb,bibcgs}.
Linear spin wave theory\cite{bibejsd} using the Heisenberg Hamiltonian
\begin{equation}
{\cal H}=-\sum_{i\mu,j\nu}J_{i\mu,j\nu}{\bf S}_{i\mu}
\cdot{\bf S}_{j\nu}-\sum_{i\mu}\epsilon_{\mu}
H_{\mu}S^z_{i\mu},
\end{equation}
satisfactorily describe the experimental
dispersion relation. The   dominant exchange constants for Cr$_2$O$_3$
are $J=-7.5$ and $-3.3$ meV between spin pairs $A$ and $B$ respectively\cite{bibejsa}  
(refer to Fig.~\ref{spin_stru}), while for $\alpha$-Fe$_2$O$_3$
spin pairs $C$ and $D$ interact most strongly with
exchange constants $J=-2.6$ and $-2.0$ meV respectively.\cite{bibejsb} 
In V$_2$O$_3$, the corundum symmetry is 
broken\cite{ltcepw,bibdbmh} at the AFI transition.
Consequently there are seven different
exchange constants between these spin pairs
[refer to Fig.\ref{spin_stru}(b)].

Dispersion relations for spin waves in the 
AFI phase of $\rm V_2O_3$ are derived in Appendix~\ref{app_afi}
following the procedure described by S\'{a}enz\cite{sw_saenz}. 
The limited data in Fig.~\ref{fig_disp} do not allow
reliable determination of individual exchange constants.
Along the $(hh0)$ direction the acoustic spin wave has a simple dispersion
relation
\begin{eqnarray}
[\hbar\omega(\bbox{\kappa})]^2&=&[H_0-4S(J_{\beta}
+J_{\delta}+2J_{\zeta})]^2 \nonumber \\
& &-[4S(J_{\beta}+J_{\delta}+2J_{\zeta})]^2\cos^2(2\pi h).
\end{eqnarray}
In combination with the previously measured value of the ordered moment
($gS=1.2$\cite{bibrmm}) our measurements of this dispersion relation 
permits the local anisotropy field, $H_0=0.13(2)$ meV, and
$J_{\beta}+J_{\delta}+2J_{\zeta}=33(2)$ meV to be 
determined with confidence.\cite{note1} To determine the individual
exchange constants, 
experiments which will measure spin waves in the AFI across the
whole Brillouin zone are now under way.

\section{The insulating spin liquid}
\label{sec_pi}

\begin{figure}[bt]
\centerline{
\psfig{file=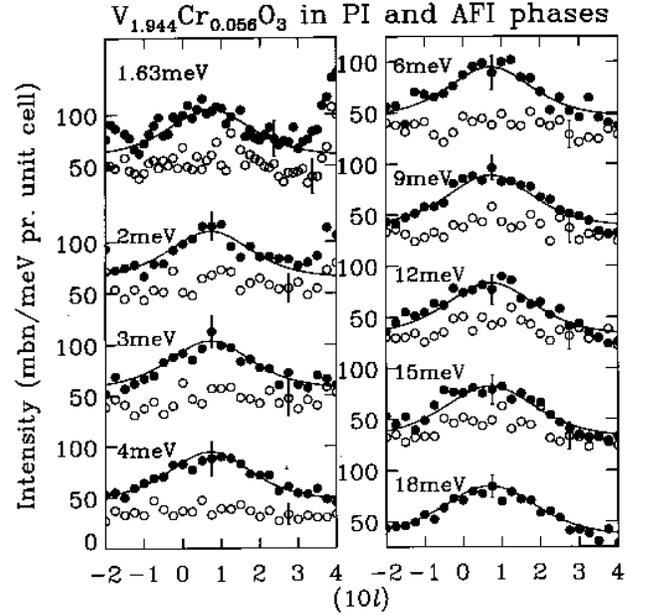,width=\columnwidth,angle=0,clip=}}
\caption{Constant energy scans along the {\bf c} direction
in the PI phase at 205~K (solid circles)
 and in the AFI phase (open circles) for (V$_{1-x}$Cr$_x$)$_2$O$_3$
(x=0.028). Measurements in the AFI phase were conducted at 170~K
for $\hbar\omega=6$-12 meV and at 160~K for other energies.
Extra intensity near the nuclear Bragg points (10$\overline{2}$) and
(104) for $\hbar\omega \leq 2$~meV is probably due to phonons. 
Statistical uncertainty in intensity is indicated by the vertical bars. 
The spectrometer configurations at BT2 were
$E_i=12.7$~meV with horizontal collimations 60'-40'-40'-60' for
$\hbar\omega=1.63$ meV, $E_f=13.7$~meV with 60'-40'-40'-60' for
$\hbar\omega=2$-15 meV, and $E_i=35$~meV with 60'-40'-80'-100' for 
$\hbar\omega=18$ meV. 
} 
\label{cr_piafi}
\end{figure}

\begin{figure}[bt]
\centerline{
\psfig{file=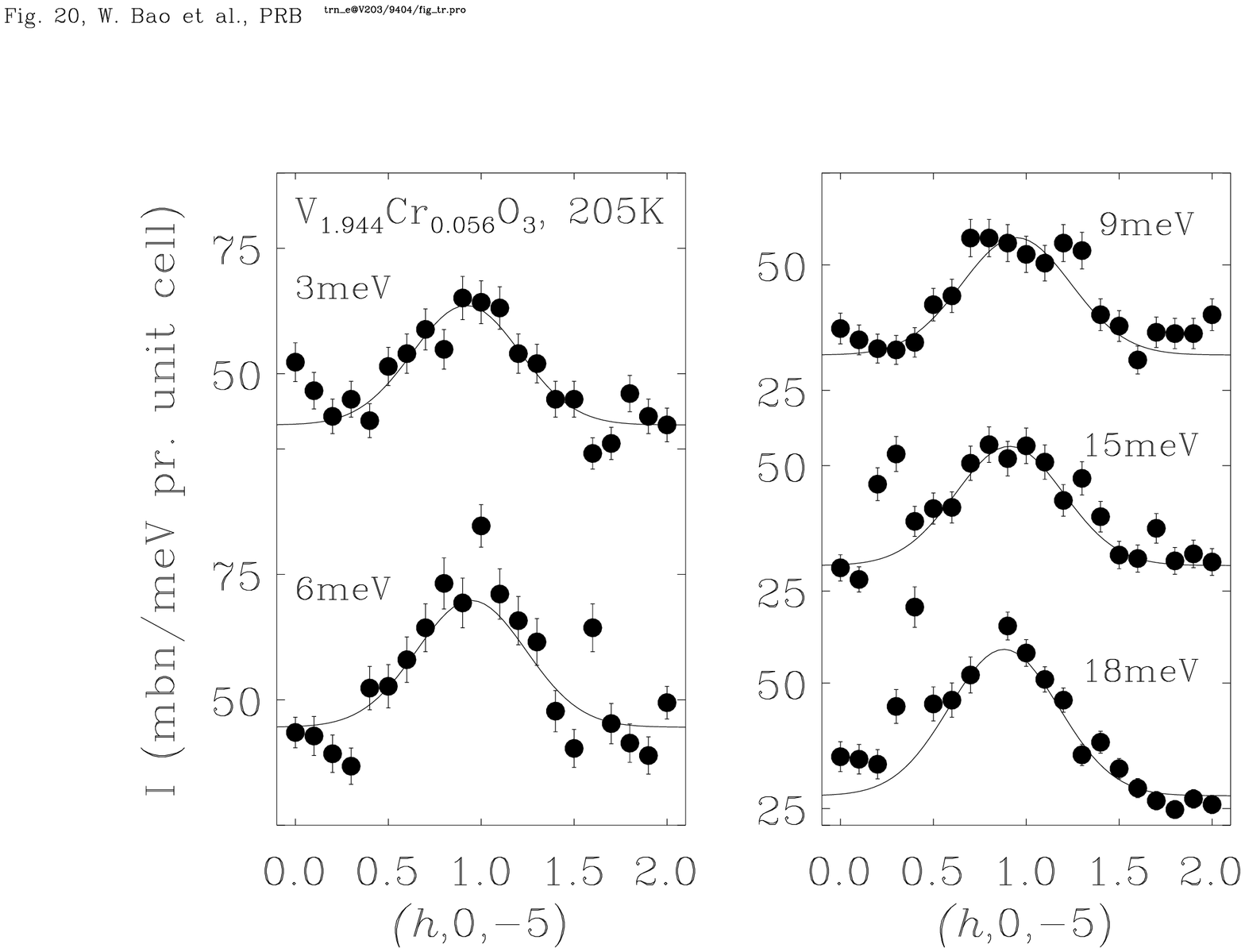,width=\columnwidth,angle=90,clip=}}
\caption{Constant energy scans along the {\bf a}$^*$ direction
in the PI phase at 205~K for (V$_{1-x}$Cr$_x$)$_2$O$_3$ (x=0.028).
There is no appreciable energy dependence of the peak width.
The spectrometer configurations at BT2 were $E_f=13.7$~meV with 
horizontal collimations 60'-40'-40'-60' for
$\hbar\omega=3$-15 meV, and $E_i=35$~meV with 60'-40'-80'-100' for 
$\hbar\omega=18$ meV. 
}
\label{cr_lng}
\end{figure}

The solid circles in Fig.~\ref{cr_piafi} represent constant energy scans 
along (10$\ell$) at 205K in the PI phase of V$_{1.944}$Cr$_{0.056}$O$_3$. 
The peak profile is not sensitive to energy transfer for 
1.6 meV$\le\hbar\omega\le$18 meV. The peaks are
very broad and the half-width-at-half-maximum (HWHM) corresponds
to a correlation length $\xi_c \sim 1.5\AA$ in the {\bf c} direction.
Scans in the basal plane (refer to Fig.~\ref{cr_lng})
also yielded broad $\hbar\omega$-insensitive peaks,
corresponding to $\xi_a \sim 2.0\AA$ in the {\bf a} direction.
These correlation lengths extend only to the nearest neighbors in
the respective directions and they are even shorter than in the PM phase
at comparable temperatures,
as can be ascertained by comparing Fig.~\ref{pure_doped} 
or Fig.~\ref{fig_paf}
to Fig.~\ref{cr_piafi}. This is a surprising result 
because the electron-hole pair damping mechanism which
exists in the metal is inhibited
by the charge excitation gap of the insulator. 

\begin{figure}[bt]
\centerline{
\psfig{file=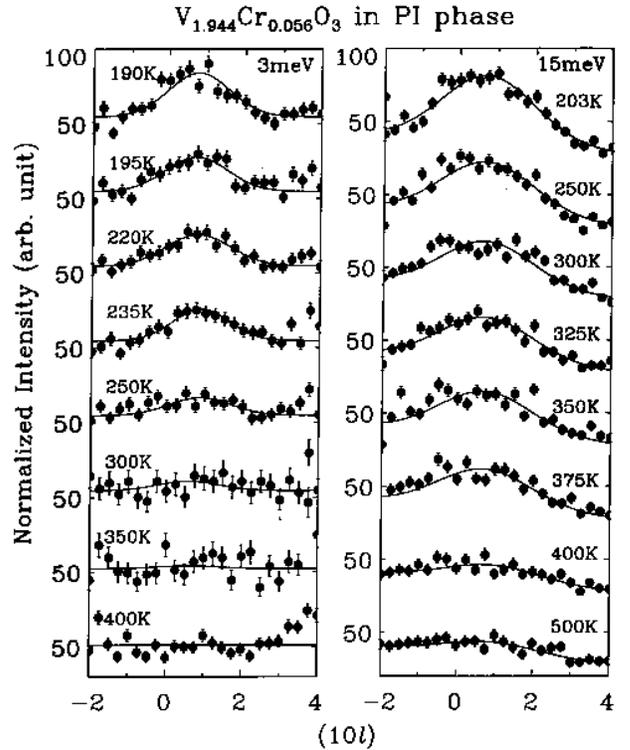,width=\columnwidth,angle=0,clip=}}
\caption{Temperature dependence of constant energy scans in the PI phase 
of V$_{1.944}$Cr$_{0.056}$O$_3$. The left panel is for $\hbar\omega=3$~meV,
and the right panel for $\hbar\omega=15$~meV. The spectrometer 
configurations at BT2 were $E_f=13.7$~meV with horizontal 
collimations 60'-40'-40'-60'. The solid line is a fit to
the product of the form
factor $|f({\bf q})|^2$ and a gaussian function.
}
\label{cr_tdep}
\end{figure}

The effect of temperature on the short-range magnetic correlations
in the PI phase is shown in Fig.~\ref{cr_tdep}. Data for two 
values of energy transfer are shown.
Clearly, {\bf q}-dependent magnetic scattering is suppressed at high
temperatures.
Even so, there is no discernible narrowing
of the peak, even as the AFI transition temperature, $T_N=180$~K, is
approached. Thus, $T_N$ does not appear to be the critical temperature for spin
fluctuations in the PI.
As for stoichiometric V$_2$O$_3$ 
(refer to Fig.~\ref{fig_paf}), the characteristic wave vector for  
magnetic correlations of Cr-doped V$_2$O$_3$ moves to 
another part of reciprocal space
in the AFI phase\cite{bao96c}
leaving a flat background as shown by open circles in Fig.~\ref{cr_piafi}.
Therefore, the phase transition to the AFI is not
a conventional antiferromagnetic transition, for which the correlation length
diverges at $T_N$ and magnetic Bragg peaks  emerge
from  peaks in the paramagnetic response function.
Instead, the spin Hamiltonian is altered by some underlying 
processes.

Not only is the location in reciprocal space
of the paramagnetic neutron scattering 
in the PI phase unusual, but also is the
$\bf q$ and $\hbar\omega$ dependence of  scattering
about this location.
For conventional insulating antiferromagnets, 
when the correlation length is limited
to the nearest neighbor spin spacing by thermal fluctuations at $T \gg J/k_B$,
there is very little spectral weight in the modulated part of the 
dynamic correlation function. In the PI phase of 
$\rm V_{1.944}Cr_{0.056}O_3$, however, constant energy cuts through 
${\cal S}({\bf q},\omega )$ show that large
spectral weight ($>1.4\mu_B$ per V, see below) is associated with a
ridge along $\hbar\omega$ which forms 
a very broad peak in $\bf q$-space. 
These experimental results suggest that the spin-spin interaction strength 
is not dwarfed by the thermal energy and that a mechanism other than thermal fluctuations therefore must be 
responsible for limiting  magnetic correlations to  nearest
neighbors. 

One possible mechanism is 
geometric frustration where conflicting
magnetic interactions prevent long range order at $T\sim J/k_B$.
Solid state chemistry, however, does not support this explanation  
because the dominant exchange interactions in Heisenberg models for
corundum antiferromagnets in general are not frustrated\cite{bibefb,bibnmkdb}.
Disorder introduced by Cr doping is also unlikely to be important, 
since for the same sample at low 
temperatures, spin waves in the AFI phase are
resolution-limited (refer to section \ref{sec_afi} and Fig.\ 3 
in Ref.\ \cite{bao96c}).
A very promising explanation, 
based on the electronic state of V$_2$O$_3$ in the single ion limit,
was recently proposed by Rice\cite{bibrice,bao96c}.
Each magnetically active
electron on a vanadium ion can exist in one of the two doubly
degenerate $3d$ orbitals.\cite{bibjbg} 
The orbital degeneracy may be the source of large orbital
contribution to the nuclear relaxation\cite{vtaki}.
Depending on the relative occupation of these
two degenerate orbitals on neighboring sites, the exchange interaction
between spins on these sites can have different 
signs\cite{bibcca}.
In the PI phase, the orbital occupation is fluctuating and disordered, 
resulting in a spin Hamiltonian with fluctuating exchange 
interactions of different signs
which induce local antiferromagnetic correlations,
while dephase longer range correlations. 
We will return to this idea in section~\ref{sec_trn}.

\subsection{Analysis of spatial correlations}

\label{spatial}
A consequence of the $\hbar \omega$-insensitive peak width in the PI phase
(Fig.~\ref{cr_piafi} and \ref{cr_lng}) is that constant energy scans directly
probe the $\bf q$-dependence of the instantaneous spin correlation function.
Because of the short correlation lengths, the spatial dependence of the
magnetic correlations can be modeled by a finite spin cluster
which includes the four types of near neighbor spin pairs: $A$, $B$, 
$C$ and $D$ [refer to Fig.~\ref{spin_stru}(a)]. 
In other words, the summation in the structure factor
\begin{equation}
{\cal S}({\bf q})\equiv \frac{(g\mu_B)^2 n}{3}
	\left( \langle {\rm S}_{\bf 0} \rangle^2 +
	\sum_{\bf R'}\cos({\bf q}\cdot {\bf R'})
	\langle {\bf S}_{\bf 0}\cdot{\bf S}_{\bf R'}\rangle 
	\right) \label{eq_clst}
\end{equation}
needs only include one $A$-type spin pair, three $B$-type spin pairs,
three $C$-type spin pairs and six $D$-type spin pairs for any
one spin, $\bf S_0$, of the $n=12$   spins in a hexagonal unit cell.

\begin{figure}[bt]
\centerline{
\psfig{file=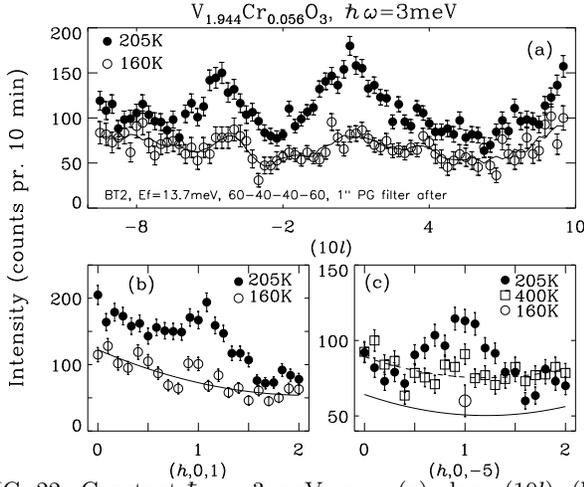,width=\columnwidth,angle=90,clip=}}
\caption{Constant-$\hbar\omega=3$~meV scans (a) along (10$l$),
(b) along ($h$,0,1), and (c) along ($h$,0,$\overline{5}$)
at 205~K in the PI phase (solid circles) and at 160~K in the AFI phase
(open circles). Squares in (c) represent data at 400~K.
(1,0,$\overline{8}$), (1,0,$\overline{2}$), (1,0,4) and
(1,0,10) are nuclear Bragg points. The experimental
configuration is indicated in frame (a). 
}
\label{pi_raw}
\end{figure}

\begin{figure}[bt]
\centerline{
\psfig{file=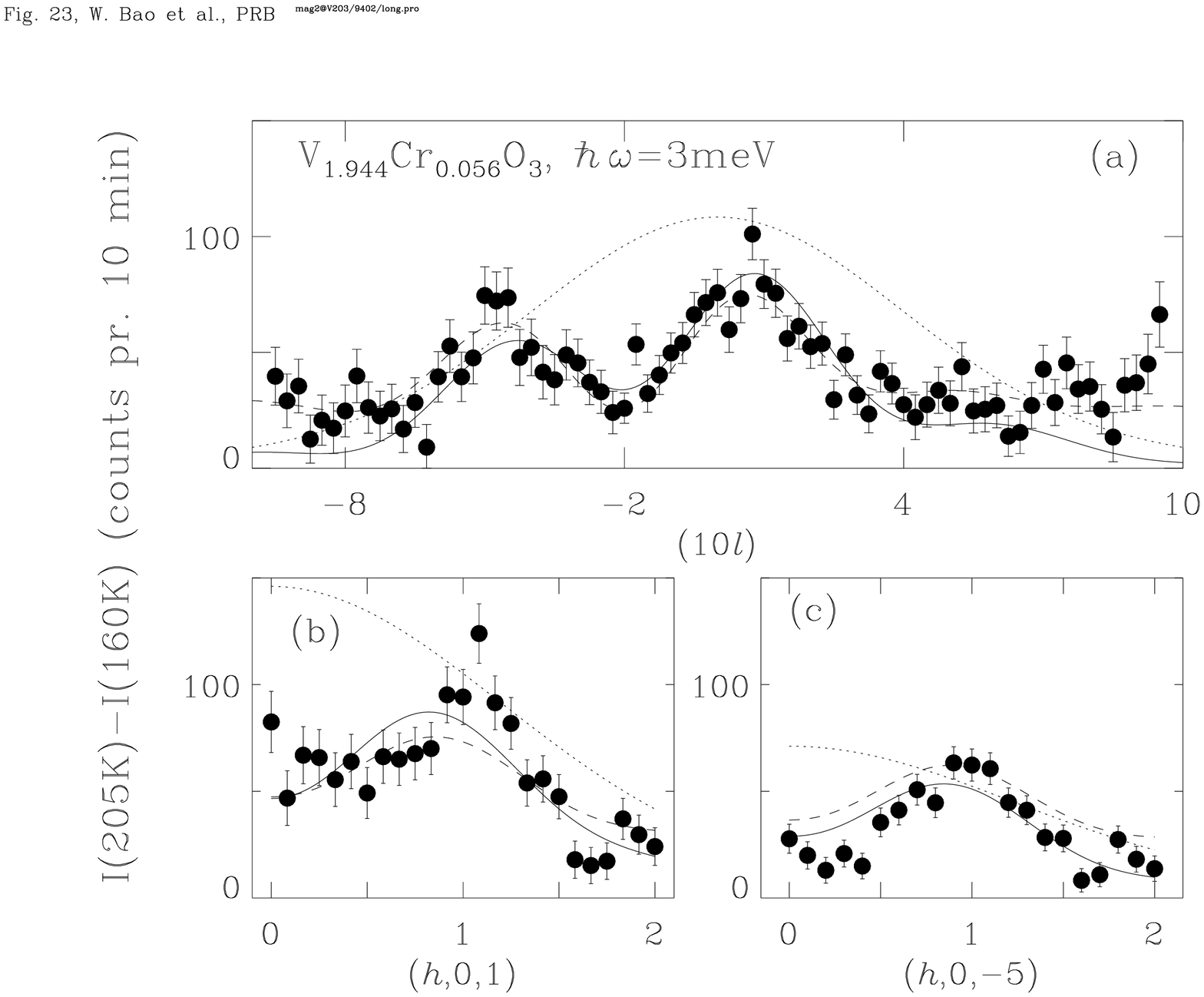,width=\columnwidth,angle=90,clip=}}
\caption{Magnetic signal at 205~K in the PI phase, 
after subtracting the smoothed
160~K intensity (solid line in Fig.~\ref{pi_raw}) as a measure
of non-magnetic backround.
The background for (c) was determined from the 400~K and 160~K data.
The dotted line represents the magnetic form factor $|f({\bf q})|^2$.
Extra intensity at large $|{\bf q}|$
is probably due to phonon scattering.
The solid line represents
$|f({\bf q})|^2 {\cal S}({\bf q})$ calculated 
with the parameters of Eq.~(\ref{eq_clstp}) while the dashed line
is a constant plus this quantity calculated with  the parameters
of Eq.~(\ref{eq_clstp_con}).}
\label{pi_mag}
\end{figure}

In Fig.~\ref{pi_raw}, an extended scan along the {\bf c} axis,
covering more than three Brillouin zones, and scans in the basal
plane intersecting the peaks near the (101) and (10$\overline{5}$)
are shown for V$_{1.944}$Cr$_{0.056}$O$_3$ at 205~K in the PI phase.
Background measured at 160K in the AFI phase is also shown.
A good approximation to the magnetic intensity at 205~K is 
obtained by subtracting 
the 160~K background and such difference data are shown in Fig.~\ref{pi_mag}.
The dotted line indicates the 
magnetic form factor for the $V^{3+}$ ion\cite{bibrewajf}, $|f({\bf q})|^2$. 
Extra intensity at large $|{\bf q}|$ is attributed
to incomplete subtraction of phonon neutron scattering.
A simultaneous fit of $|f({\bf q})|^2 {\cal S}({\bf q})$ to the magnetic
intensity from these 3~meV scans is shown as solid lines in 
Fig.~\ref{pi_mag}. As can be seen in the figure 
this spin cluster model is consistent with the data
over the large part of
{\bf q}-space where the magnetic form factor is appreciable.

The near-neighbor spin correlations derived from the best fit 
($\chi^2=1.9$)
are:
\begin{mathletters}
\label{eq_clstp}
\begin{eqnarray}
\langle {\bf S}_{\bf 0}\cdot {\bf S}_{[0,0,1/6+\delta]}\rangle^A 
/\langle {\rm S} \rangle^2 &=& 0.08(9), \\
\langle {\bf S}_{\bf 0}\cdot {\bf S}_{[1/3,2/3,\overline{\delta}]}\rangle^B 
/\langle {\rm S} \rangle^2 &=&-0.10(4),\\
\langle {\bf S}_{\bf 0}\cdot {\bf S}_{[2/3,1/3,\delta-1/6]}\rangle^C
/\langle {\rm S} \rangle^2 & =&0.04(4),\\
\langle {\bf S}_{\bf 0}\cdot {\bf S}_{[2/3,1/3,1/6]}\rangle^D
/\langle {\rm S} \rangle^2 &=&-0.06(2), 
\end{eqnarray}
\end{mathletters}
where $\delta=0.026$ and superscripts $A$-$D$ refer to spin pairs
in Fig.~\ref{spin_stru}(a). 
Now, it is likely that warming yields an additional non-magnetic background,
most probably due to phonon and multiphonon scattering responsible
for the upturn at high $|l|$ in Fig.~\ref{pi_raw}(a).
We account for this rising background by incorporating an
additive constant in our fits, as we have done previously\cite{bao95d,bao96c}. 
The best least squared fit ($\chi^2=1.2$)
is obtained with magnetic parameters\cite{bao95d,bao96c}
\begin{mathletters}
\label{eq_clstp_con}
\begin{eqnarray}
 \langle {\bf S}_{\bf 0}\cdot {\bf S}_{[0,0,1/6+\delta]}\rangle^A
/\langle {\rm S} \rangle^2 &=& 0.6(3), \\
 \langle {\bf S}_{\bf 0}\cdot {\bf S}_{[1/3,2/3,\overline{\delta}]}\rangle^B  
/\langle {\rm S} \rangle^2 &=&-0.19(8), \\
 \langle {\bf S}_{\bf 0}\cdot {\bf S}_{[2/3,1/3,\delta-1/6]}\rangle^C
/\langle {\rm S} \rangle^2 & =&0.18(8), \\
 \langle {\bf S}_{\bf 0}\cdot {\bf S}_{[2/3,1/3,1/6]}\rangle^D
/\langle {\rm S} \rangle^2 &=&-0.09(3),  
\end{eqnarray}
\end{mathletters}
and a constant background of 25 counts in 10 min.

The physical picture which emerges from both fits is that in the PI phase,
the nearest neighbor spin pairs in the basal plane, which contains a
puckered honeycomb lattice for V, are more likely to be anti-parallel
to each other, while the nearest neighbor
spin pairs in the {\bf c} direction are more likely to be parallel
to each other. While the signs of correlations for different
pairs are clear, the specific values of the spin pair correlations
have large statistical error bars and possibly systematic errors
associated with uncertainty in the  background determination. 
If this dynamic short-range spin correlations were allowed to 
develop into a long-range order, the spin correlations would be
\begin{eqnarray}
 \langle {\bf S}_{\bf 0}\cdot {\bf S}_{[0,0,1/6+\delta]}\rangle^A
/\langle {\rm S} \rangle^2 &=& 1, \nonumber \\
 \langle {\bf S}_{\bf 0}\cdot {\bf S}_{[1/3,2/3,\overline{\delta}]}\rangle^B  
/\langle {\rm S} \rangle^2 &=&-1, \nonumber \\
 \langle {\bf S}_{\bf 0}\cdot {\bf S}_{[2/3,1/3,\delta-1/6]}\rangle^C
/\langle {\rm S} \rangle^2 & =&1, \nonumber \\
 \langle {\bf S}_{\bf 0}\cdot {\bf S}_{[2/3,1/3,1/6]}\rangle^D
/\langle {\rm S} \rangle^2 &=&-1. \nonumber  
\end{eqnarray}
This corresponds to the type I collinear antiferromagnetic structure for
corundum\cite{bibefb}.

\begin{figure}[bt]
\centerline{
\psfig{file=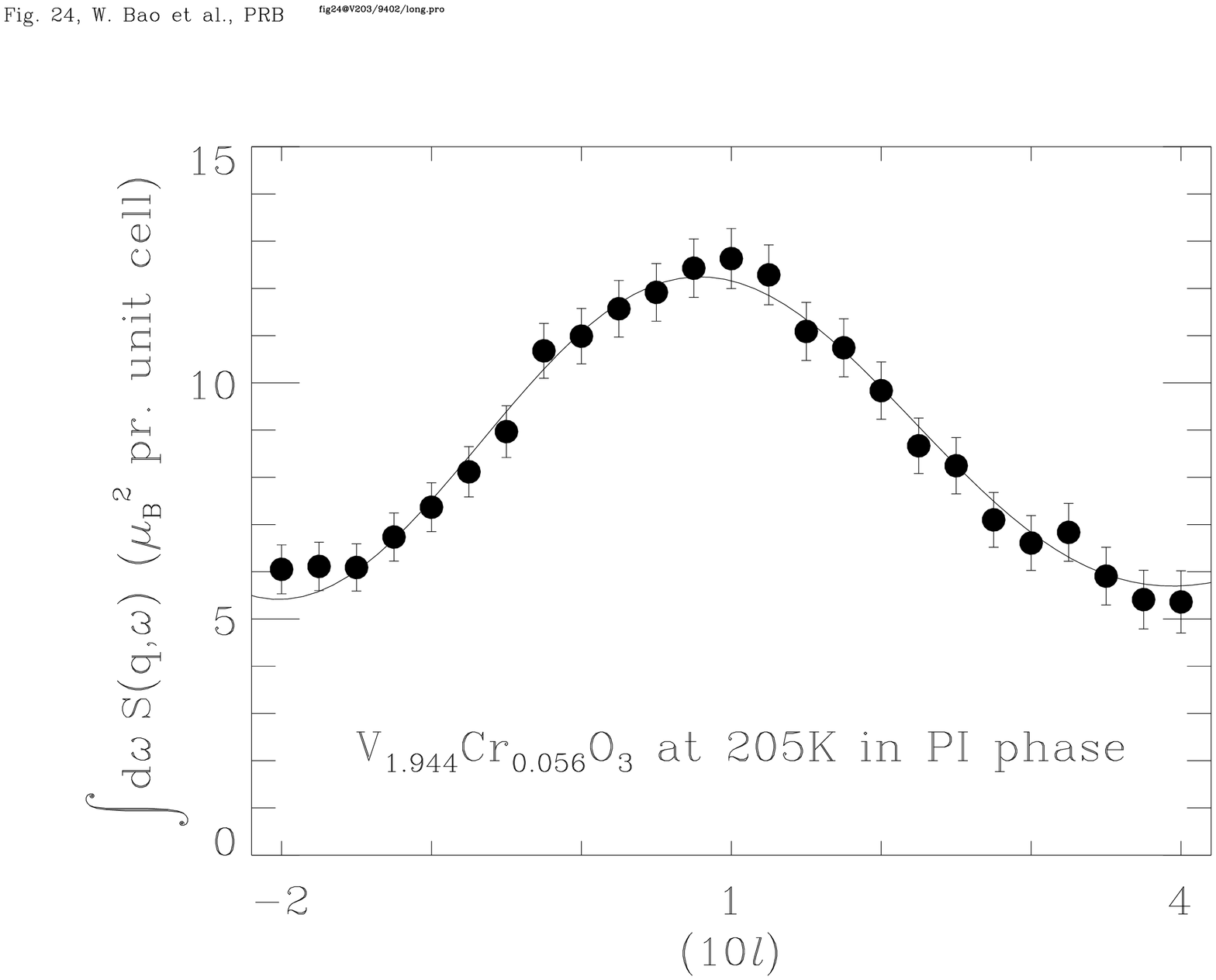,width=.8\columnwidth,angle=90,clip=}}
\caption{Energy integrated intensity, up to 18 meV, as a function 
of ${\bf q}=(10\ell )$
at 205K in the PI phase for $\rm V_{1.944}Cr_{0.056}O_3$.
The solid line is $12  {\cal S}({\bf q})$ based on Eq~(\ref{eq_clst}), the 
parameters of Eq.~(\ref{eq_clstp}), and 
$g^2 {\rm S}^2 /3=0.80^2$. }
\label{pi_int}
\end{figure}

In Fig.~\ref{pi_int}, the magnetic intensities
of Fig.~\ref{cr_piafi} from $\hbar\omega=1.63$~meV
to 18~meV have been integrated, to yield the 
quasi-instantaneous correlation function
\begin{eqnarray}
	\hbar\int d\omega\, {\cal S}({\bf q},\omega)&\approx &
	\frac{(g\mu_B)^2}{3 N}
	\sum_{{\bf R}{\bf R}'} e^{i{\bf q}\cdot({\bf R}'-{\bf R})}
	\langle {\bf S}_{\bf R}(\delta t)\cdot 
	{\bf S}_{{\bf R}'}(0)\rangle \nonumber \\
& \approx &{\cal S}({\bf q}),
\end{eqnarray}
where $|\delta t| <\hbar/18$meV$=3.7\times 10^{-14}$sec. 
The solid line in the figure is
${\cal S}({\bf q})$, calculated from Eq.\ (\ref{eq_clst}),
(\ref{eq_clstp}) using 
$(g\mu_B)^2\langle {\rm S}\rangle^2/3=(0.80\mu_B)^2$.
Thus, at 205K, the fluctuating moment
involved in the short range correlations, up to $\hbar\omega=18$~meV,
is $1.4(2)\mu_B$ per V. This is as large as the ordered moment
of 1.2$\mu_B$ per V in the AFI phase. 

\subsection{Energy dependence}
\label{edep}

\begin{figure}[bt]
\centerline{
\psfig{file=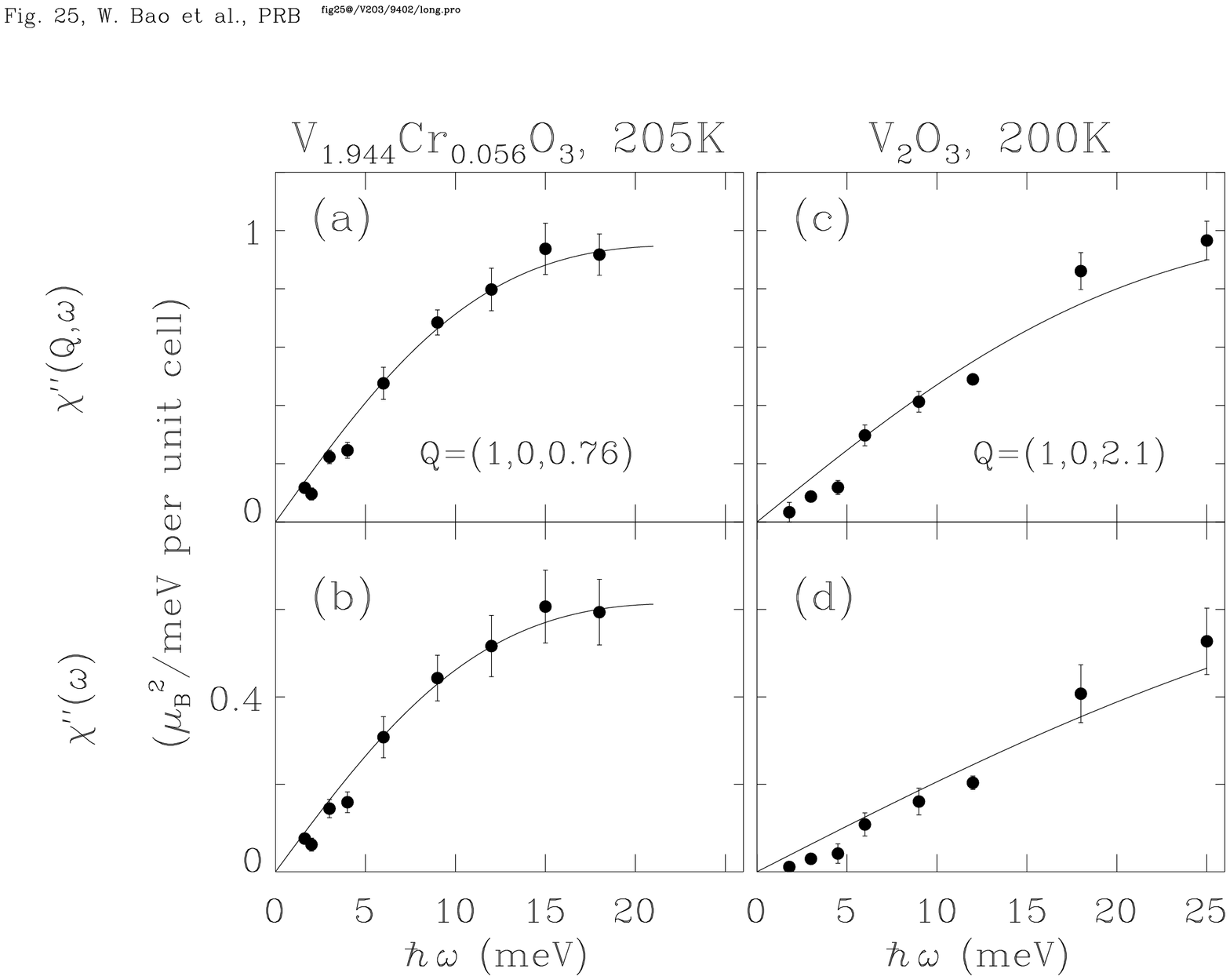,width=\columnwidth,angle=90,clip=}}
\caption{(a) The dynamic susceptibility at {\bf Q} specified in the
figure, and (b) the local dynamic susceptibility as a function of
energy for V$_{1.944}$Cr$_{0.056}$O$_3$ at 205K in the PI phase.
The corresponding quantities for stoichiometric V$_{2}$O$_3$ at 200K
in the PM phase are shown in (c) and (d).
}
\label{fig_cmpe}
\end{figure}

Fig.~\ref{fig_cmpe}(a) shows the energy dependence of  $\chi''({\bf Q},\omega)$ at
${\bf Q}=(1,0,0.76)$ for V$_{1.944}$Cr$_{0.056}$O$_3$  at 205 K. The data 
were obtained from the peak intensities of the
constant-energy scans in Fig.~\ref{cr_piafi}. 
There is a remarkable similarity between these data and the peak intensities
for metallic pure $\rm V_2O_3$ shown in Fig.~\ref{fig_cmpe}(c) despite the 
difference in $S(\bf q)$ for these two samples 
(see, e.g., Fig.~\ref{pure_doped} and \ref{cr_piafi}).

The spectrum can be modeled
by a Lorentzian
\begin{equation}
\chi''({\bf q},\omega)=\frac{\chi({\bf q})}{\Gamma_{\bf q}}\, 
\frac{\hbar\omega}{1+(\hbar\omega/\Gamma_{\bf q})^2},\label{eq_pif}
\end{equation}
where $\chi({\bf q})$ is a generalized 
{\bf q}-dependent static susceptibility and
$\Gamma_{\bf q}$ is the relaxation energy.
The  data in Fig.~\ref{fig_cmpe}(a) only allow
us to place a lower limit of 18~meV on $\Gamma_{\bf Q}$
at $T=205$ K. The initial slope
\begin{equation} 
\Pi ({\bf Q} )\equiv \lim_{\omega\rightarrow 0}
\chi''({\bf Q},\omega)/\hbar\omega 
=\chi({\bf Q})/\Gamma_{\bf Q}
\label{EqPi}
\end{equation}
is however reliably determined to be 
$\Pi ({\bf Q} )= 0.075(4)\mu_B^2/$meV$^2$ per unit cell.

The energy dependence of the local magnetic susceptibility
\begin{equation}
\chi''(\omega)\equiv \int d^3{\bf q} 
\chi''({\bf q},\omega)/\int d^3{\bf q},
\end{equation}
for the PI and PM are shown in Fig.~\ref{fig_cmpe}(b) and (d).
For the insulating sample these data were derived from the Gaussian fits 
shown in Figs.~\ref{cr_piafi} and \ref{cr_lng} while we used the fits shown in 
Fig.~\ref{fig_paf} for the metallic sample. 
Again we observe that 
for $T\approx 200$ K the magnetic spectral weight for $\hbar\omega < 25$ meV
is substantially larger in the PI phase than in the PM phase.

\subsection{Comparison to bulk susceptibility data}

As a check on the reliability of our normalization procedures,
we compare the normalized dynamic susceptibilities at $T=205$ K 
discussed above 
to the conventional bulk susceptibility, $\chi\equiv\chi'({\bf 0},0)
=\chi({\bf q= 0})+\chi'({\bf 0},\infty)$
for $\rm V_{1.944}Cr_{0.056}O_3$.

From Eq.\ (\ref{eq_pif}) and the Kramers-Kronig relation, 
as long as we are in the classic regime,
$\Gamma_{\bf q} < k_B T$,
\begin{equation}
{\cal S}({\bf q})\equiv\hbar\int d\omega\, \chi''({\bf q},\omega)\,
\langle n(\omega)+1\rangle\approx k_B T \chi({\bf q}).
\end{equation}
Therefore, at a given temperature,
\begin{equation}
\chi({\bf 0}) \approx \frac{{\cal S}({\bf 0})}{{\cal S}({\bf Q})}\,
\chi({\bf Q}) 
= \frac{{\cal S}({\bf 0})}{{\cal S}({\bf Q})}\,
\Gamma_{\bf Q}\,\Pi ({\bf Q} ).
\end{equation}
Using parameters in (\ref{eq_clstp}) or (\ref{eq_clstp_con})
for our finite cluster model, we find
$ S({\bf Q})/S({\bf 0})\approx 3.0$ or 2.2.
Thus, assuming that $\Gamma_{\bf Q} \lesssim k_B T$, we can estimate the
bulk susceptibility $\chi({\bf 0})$ from the
parameters quoted in section \ref{edep}.
The result is
$\chi\geq 0.5$-$0.7\mu_B^2$/meV per unit cell $=1.1$-$1.6\times10^{-3}$
emu/mole V for V$_{1.944}$Cr$_{0.056}$O$_3$ at $T=205$ K.
This number should be compared to the measured bulk susceptibility
which is  
$3.6\times10^{-3}$ emu/mole V at 205K.\cite{bibdbme}  

\subsection{Temperature dependence}

\begin{figure}[bt]
\centerline{
\psfig{file=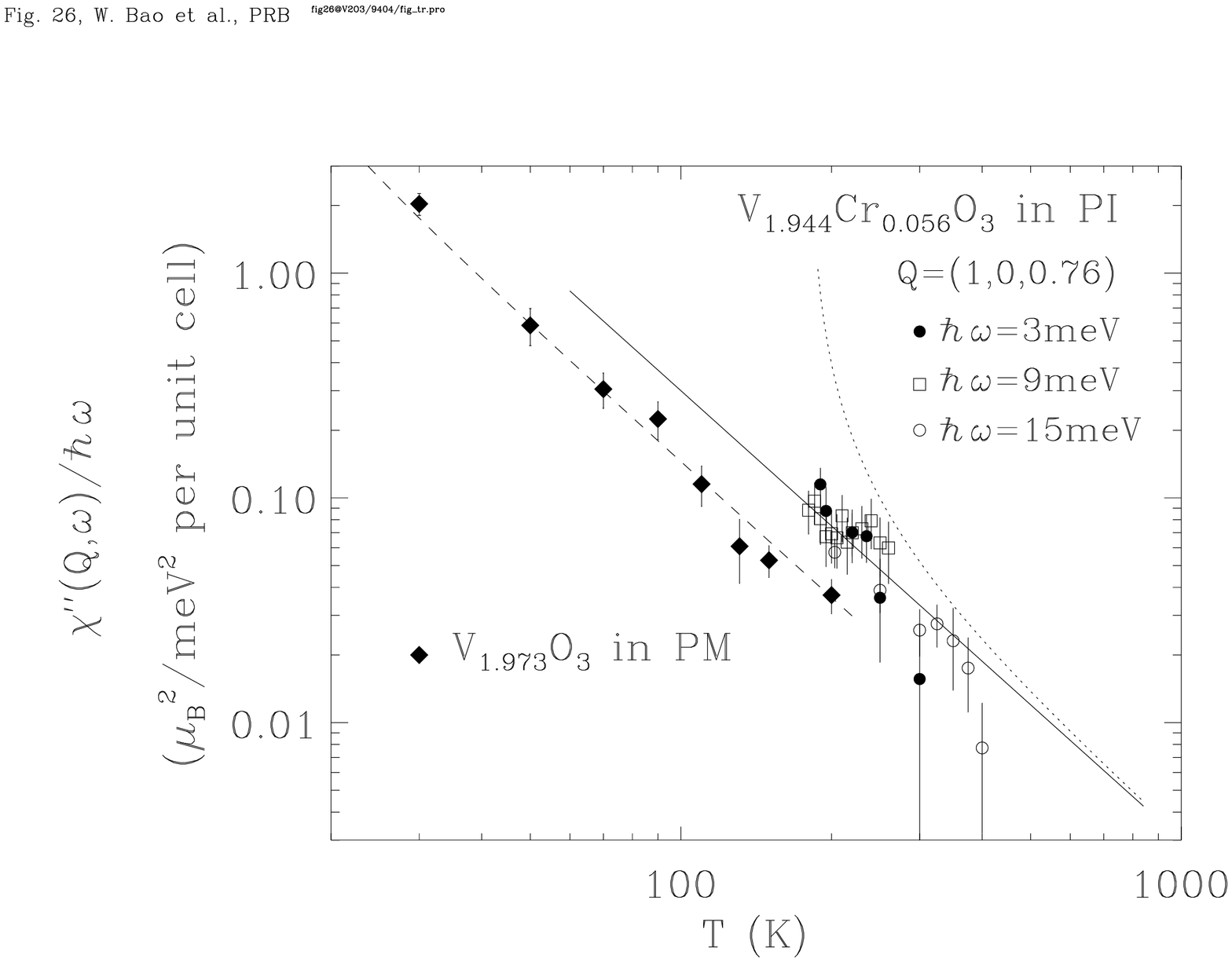,width=\columnwidth,angle=90,clip=}}
\caption{Temperature variation of $\chi''({\bf Q},\omega)/\hbar\omega$
at ${\bf Q}=(1,0,0.76)$
for V$_{1.944}$Cr$_{0.056}$O$_3$ in the PI phase, with
$\hbar\omega$ used in the measurement indicated in the figure. 
The corresponding quantity at the low energy limit, 
$\chi_{\bf Q}/\gamma_{A}\kappa^2$, for V$_{1.973}$O$_3$ in PM
(diamonds) is included for reference. 
The solid line is proportional to $T^{-2}$, 
the dotted line is proportional to
$(T^2-(180\ K)^2)^{-1}$, and the dashed line is proportional to
$(T^2-(9\ K)^2)^{-1}$.}
\label{cr_tdepf}
\end{figure}

The temperature dependence of $\chi''({\bf Q},\omega)/\hbar\omega$ 
is shown in Fig.~\ref{cr_tdepf} for V$_{1.944}$Cr$_{0.056}$O$_3$. 
The 3 meV and 15 meV data are from Fig.~\ref{cr_tdep} and 
the 9 meV data from Fig.~4(c) of Ref.~\cite{bao96c}.
The values of $\chi''({\bf Q},\omega)/\hbar\omega$ from different
$\hbar\omega=3$-15 meV are close to each other down to $T_N=180$~K. 
This indicates that $\chi''({\bf Q},\omega )\propto\hbar\omega $
in this energy range and that $\Gamma_{\bf Q}$ therefore exceeds
15 meV even close to the AFI transition (refer to Eq.~(\ref{eq_pif})).
This also means that the ratios plotted are a measure of $\Pi ({\bf Q} )$ [Eq.~(\ref{EqPi})].

In a second-order antiferromagnetic transition, $\chi({\bf q})\propto
(T+T_N^0)^{-1}$ and $\Gamma_{\bf Q} \propto T-T_N^0$ in the Gaussian 
approximation\cite{Ma_book}, therefore, 
$\Pi ({\bf Q} )\propto
(T^2-T_N^{0\, 2})^{-1}$ scales as $T^{-2}$ for $T\gg T_N^0$.
As expected, experimental data for metallic V$_{1.973}$O$_3$ 
follow this relation
for $T\gg T_N=9$ K
(refer to diamonds in Fig.~\ref{cr_tdepf}). 
Interestingly, the data for the temperature dependence of 
$\Pi ({\bf Q} )$ in V$_{1.944}$Cr$_{0.056}$O$_3$, 
are proportional to $T^{-2}$ (refer to the solid line)
all the way down to $T_N=180$~K.
Therefore, the spin system in the PI phase at temperatures
even slightly above the PI-AFI phase transition appears to be 
far away from the criticality associated with the real $T_N$. 
On the other hand, $T_N^0$ is remarkably close to
N\'{e}el temperature of the metallic AFM found for
hole-doped samples. In other words, the PI fluctuations are those which 
one would associate with nearly the same quantum critical
point as found for the PM and AFM phases.

\section{Discussion and Conclusions}
\label{sec_trn}

In the previous sections, the magnetic properties 
of V$_2$O$_3$ in each of its
phases were described separately. 
Interest in this material, however, is driven by attempts to understand
the Mott-Hubbard transition. Here, we will discuss the implications of
our magnetic neutron scattering experiments for the interpretation
of the transitions between the PI, AFI, PM, and AFM phases.

The discovery of antiferromagnetic order in metallic
V$_2$O$_3$ through the $^{57}$Fe M\"{o}ssbauer 
effect\cite{bibyukk}, prompted much theoretical
work on antiferromagnetism in strongly correlated Fermi liquids.
Phase diagrams including a metallic antiferromagnetic phase for the Mott
system were subsequently published by several 
authors\cite{mthbcyrl,scrmorv,bibjsad,inftymitb}.  
However, for many of
these studies, a Heisenberg antiferromagnetic state like that
in the insulator was put into the metallic state {\em by hand} 
and local moment spin dynamics in this phase was implicit.
Even though the effective mass of electrons in metallic V$_2$O$_3$
is strongly enhanced, i.e., the Fermi liquid is nearly localized
near the metal-insulator transition, we showed in the previous sections that
the antiferromagnetism in metallic V$_2$O$_3$ is completely different from 
the antiferromagnetism in the AFI. While Heisenberg models
of exchange coupled unpaired spins located at V ions
can account for magnetic phenomena in the AFI phase, antiferromagnetism 
in the metal is controlled fundamentally by the Fermi sea,\cite{bao93,bao96a,sdw_wga}.
Specifically the small-moment static SDW order results from a nesting 
or near-nesting Fermi surface, and the broad bandwidth spin excitation
spectrum is reminiscent of an electron-hole pair continuum.
In short, our experiments show that the metal-insulator transition 
separates two qualitatively different antiferromagnetic states, 
with localized moments in the AFI
and itinerant moments in the AFM phase.

With the large number of spin fluctuation modes in the Stoner continuum,
spin fluctuations in an itinerant antiferromagnet have a stronger
renormalizing effect on equilibrium magnetic properties than spin waves
have in an insulating antiferromagnet. This makes magnetism at finite
temperatures for itinerant magnets much more complex than for 
localized moment magnets.
Near the quantum critical point in the small moment limit,
scaling relations can be derived based on general 
physical arguments\cite{ja_hertz,2dheis,2dheiqc,isfandy}.
The scaling functions and relations between non-universal
constants can be further calculated by the SCR 
theory\cite{scrmor,scrmorc,scrlnzr,scrmnsid}.
We found remarkable {\em quantitative} consistence between 
experimental results for metallic V$_{2-y}$O$_3$
and the theory.

The source of an itinerant antiferromagnetic instability,
which is characterized by singularity of generalized magnetic 
susceptibility at finite {\bf q},
could be a divergent Lindhard function 
due to nesting Fermi surface\cite{sdw_lom,sdw_paf}, a nearly
nesting Fermi surface in combination with the Coulomb 
interactions\cite{sdw_paf,hubcyrt}, or solely
the strong Coulomb interactions.
The last possibility is still under theoretic 
investigations.\cite{sq2ds,pwa_97} 
The nesting scenario has been extremely successful in describing the
classic itinerant antiferromagnet, Cr.\cite{Cr_rev}
We find it also provides a comprehensive framework
for metallic $\rm V_2O_3$\cite{bao93,bao96a}.
In contrast to Cr, however, only a small area of the Fermi surface
is involved in the SDW\cite{bao93}, and strong correlations in
metallic $\rm V_2O_3$ require only near nesting of the Fermi
surface for the long-range antiferromagnetic state to occur.

 Three kinds of metal-insulator transitions occur in the V$_2$O$_3$ system,
the PM-PI transition, the PM-AFI transition, and the T$\rightarrow$0
AFM-AFI transition.
Whether they should be called Mott transitions has been controversial and,
to some extent,
is confused by the evolving meaning of the Mott transition.
While previously noticed by others, Mott\cite{bibnfmb} was responsible
for bringing attention to the possibility of localization 
of electrons in a narrow band through electronic 
correlations, which is neglected in conventional band theories of solids. 
Mott argued for a first-order metal-insulator transition
when the electron density is reduced below some critical value.
Not much detail about the electronic processes taking place 
at the transition were  known 
at that time and since the ``exceptional'' insulators 
like NiO\cite{niocb}, which motivated Mott's
research and are today called Mott and charge-transfer insulators, 
were known experimentally to be antiferromagnets, 
it was unclear whether there was any new physics apart from magnetism.
Specifically, if these insulators are always antiferromagnetic, 
a doubling of the unit cell will open a gap at the Fermi level 
as shown by Slater\cite{slateraf}. However, Hubbard\cite{bibjhc}
demonstrated in 1964 that a metal-insulator
transition can be produced by strong Coulomb correlations without
antiferromagnetic ordering. Brinkman and Rice\cite{bibwfbrb} in 1970 outlined
a distinct type of Fermi liquid on the metallic side of the Mott-Hubbard
metal-insulator transition. Strong electron 
correlations in the PM phase of V$_2$O$_3$ were revealed 
through its Brinkman-Rice-like behavior\cite{bibdbmm} and the PM-PI
transition was discovered as a possible experimental realization of
the Mott-Hubbard transition by McWhan, Rice and Remeika\cite{bibdbmb}.
This probably stands as the first explicitly identified Mott-Hubbard
transition in a real material.

The difference between the paramagnetic metal and the paramagnetic
insulator in (V$_{1-x}$Cr$_x$)$_2$O$_3$ 
disappears at the critical point near 400~K
(or the critical line in the composition-P-T phase space\cite{bibdbmd}). 
A continuous crossover between 
paramagnetic spin fluctuations in the metal and the insulator thus
is expected on a path above the critical point.
By inspecting the evolution of magnetic correlations with temperature and
doping in the metal 
(refer to Figs.~\ref{afm_res}, \ref{pure_doped} and 
\ref{fig_paf}) and in the insulator (refer to Figs.~\ref{cr_piafi} 
and \ref{cr_tdep}), such a magnetic crossover from metal to insulator
can be conceived as follows: The pair of incommensurate peaks
in the metal broaden in $\bf q$ upon approaching the critical
point until they finally merge into the single broad peak 
which we observe in  the PI. The blurring of the incommensurate peaks
which is a measure of Fermi surface dimension on the nesting parts
parallels the losing of metallic coherence among electrons.

Although Hubbard as well as Brinkman and Rice have shown
that a Mott transition does not require an antiferromagnetic transition,
antiferromagnetic order does promote the insulating state.
This is apparent because of the enlarged insulating phase space 
associated with the AFI in various studies 
based on the one-band Hubbard 
model\cite{mthbcyrl,scrmorv,inftymitb}.
The resemblance of the theoretical 
phase diagrams to the experimental phase diagram
of V$_2$O$_3$ gives merit to Slater's idea of an insulator
induced by antiferromagnetism. 

However, an important difference exists
between these one-band phase diagrams and the V$_2$O$_3$ phase diagram.
The magnetic transition in the theories is predicted to be of
second order while the experimental AFI transition in V$_2$O$_3$ 
is first order [refer, e.g., to Fig.~\ref{P_afi}(a)]. This difference
is not trivial such as being merely a magnetostriction effect.
Beneath it is a fundamental difference in temperature dependence of
spatial spin correlations\cite{bao96c}.
It is well known\cite{hubpwa} that a one-band Hubbard model 
reduces to a Heisenberg
model with exchange $J\sim t^2/U$ in the insulating limit $U/t\gg 1$.
For temperatures lower than the insulating gap $\sim U/k_B$,
the low energy physics is contained in the Heisenberg model with
a second order antiferromagnetic transition at $T_N\sim J/k_B$.
The characteristic magnetic wave vector remains the same for
$T>T_N$ and $T<T_N$.
In V$_2$O$_3$, dynamic spin fluctuations in  the corundum PM and 
PI phases peak at a magnetic wave vector along the {\bf c}$^*$ axis.
These magnetic correlations abruptly vanish and are replaced 
by  (1/2,1/2,0)-type long range order and the associated spin waves 
below the first-order AFI transition 
(see Ref.\ [\cite{bao96c}] for details; see also 
Figs.~\ref{fig_paf} and \ref{cr_piafi} in this paper).
This switching between spin correlations with different magnetic wave vectors
indicates that the antiferromagnetic transition is not a
common order-disorder phase transition, as is found, for example, in the
parent compounds of  high-T$_C$ superconductors, but that instead the
spin Hamiltonian is somehow modified at the first-order 
transition to the AFI phase.
These results find no comprehensive explanation in one-band Hubbard models.

We mentioned in section~\ref{sec_pi} that the electronic state of V$_2$O$_3$
in the single-ion limit can be approximated by two degenerate $d$-orbitals 
at each V ion site which are  filled by a single electron.
New physics in Hubbard models with degenerate bands was discussed by several 
authors more than twenty years ago\cite{db_kkh,db_clc,db_clcr,bibcca}.
It gains renewed interest in light of recent experimental 
progress\cite{bibrice,bibpen,biboles}.
Depending on which one of the degenerate orbitals is occupied at
a given site and at its neighboring sites, the exchange interaction
between the nearest neighbor spins is different and can even change
sign. Thus, besides the spin degrees of freedom, a new orbital degree 
of freedom enters the low energy physics. 
The spin and orbital degrees of freedom are strongly coupled, 
and the effective spin Hamiltonian depends sensitively on 
orders in the orbital degrees of freedom.\cite{bibrice,bibcca}
Switching between spin correlations with different magnetic wave vectors
was also recently observed in manganite at an orbital/charge
ordering transition.\cite{bao96b}

The anomalously short-range spin fluctuations in the PI and the abrupt
switching of magnetic correlations at the AFI transition can be
understood, as first pointed out by Rice\cite{bibrice},
if the primary order parameter for the AFI transition involves  
orbital degrees of freedom.
At low temperatures, order in orbital occupations breaks the three-fold
corundum symmetry. Pairs with different orbitals at neighboring sites
produce different bond strengths and different exchange interactions,
resulting in the structural distortion and
the antiferromagnetic structure in Fig.~\ref{spin_stru} (b) which has
two antiferromagnetic bonds and one ferromagnetic bond in the basal 
plane.\cite{bibcca}
In the PI phase the orbital occupation fluctuates between
the two states of the doublet, thus restoring the
corundum symmetry and producing a different
spin Hamiltonian with random and fluctuating
exchange interactions. With this kind of spin Hamiltonian,
magnetic correlations necessarily are 
limited to the nearest neighbors.
The fact that the magnetic transition is strongly
first order indicates that the {\em intrinsic}
N\'{e}el temperature, $T_N^{AFI}$, of the 
spin Hamiltonian associated with the AFI phase is 
larger than the orbital ordering temperature $T_O$.
Thus, when orbitals order and the spin Hamiltonian changes
from the one associated with the PI phase,
which has a $T_N^{PI} < T_O$, to the one
associated with the AFI phase, magnetic moment reaches from zero
directly to a value appropriate for $T=T_O\ll T_N^{AFI}$. 

In conclusion, we have shown that 
the metal-insulator transition fundamentally
changes the nature of antiferromagnetism in the V$_2$O$_3$ system.
Metallic V$_2$O$_3$ is a prototypical strongly correlated small-moment 
itinerant antiferromagnet, involving a nesting instability 
on a small part of Fermi surface. 
We also have presented strong indications that orbital 
degeneracy plays a crucial
role in the magnetism and phase transitions of this material. 
Specifically it appears 
that these orbital degrees of freedom
give rise to a novel fluctuating paramagnetic insulator, 
and a first order transition to an antiferromagnetic phase with a
spin structure which is unrelated to the paramagnetic short range order.
At the same time, the fluctuations in the paramagnetic
insulator and paramagnetic metal are so similar that they suggest
that the electron-hole pair excitations associated with 
Fermi surface nesting in the metal are transformed into orbital
fluctuations in the insulator.

\acknowledgments

We gratefully acknowledge discussions and communications with
T. M. Rice, S. K. Sinha, Q. M. Si, G. Sawatzky, 
L. F. Mattheiss, G. Kotliar, C. Castellani, A. Fujimori,
M. Takigawa, A. J. Millis,
R. J. Birgeneau, G. Shirane, S. M. Shapiro, C. Varma, and
T. Moriya. W. B. would like to thank
R. W. Erwin, J. W. Lynn, D. A. Neumann, J. J. Rush for their hospitality
during experiments at NIST.
W. B. and C. B. thank C. L. Chien and Q. Xiao for using
their laboratory for sample preparations.
Work at JHU was supported by the NSF through DMR-9453362,
at BNL by DOE under Contract No DE-AC02-98CH10886, at ORNL
by DOE under Contract No DE-AC05-96OR22464, at Purdue by
MISCON Grant No DE-FG02-90ER45427, and at U. Chicago
by the NSF through DMR-9507873.

\appendix
\section{Spin waves in AFI} \label{app_afi}
Spin wave excitations from a {\em collinear} spin arrangement
were studied by S\'{a}enz\cite{sw_saenz}. For a material with $n$
magnetic ions in a {\em primitive} magnetic unit cell, 
the positions of these ions are described by
\begin{equation}
{\bf R}_{i\mu}= {\bf R}_{i}+{\bf r}_{\mu},
\end{equation}
where ${\bf R}_i$ is the location of the $i$th unit cell
(i=1,2,...,N), and
${\bf r}_{\mu}$ is the position of the $\mu$th ion inside
the cell ($\mu=1,2,...,n$). The vector spin operator for the
ion denoted by \{i$\mu$\} is ${\bf S}_{i\mu}$ and the spin 
quantum number is $S_{\mu}$. Choosing $\hat{z}$ to be parallel
to the collinear direction, the spin orientations are described by
$\epsilon_\mu=\pm 1$ for up or down spins in a unit cell. 
A diagonal $n\times n$ matrix, {\sf E}, with elements $\epsilon_{\mu}
\delta_{\mu\nu}$ is introduced for later use.

The spin Hamiltonian is given by
\begin{equation}
{\cal H}=-\sum_{i\mu,j\nu}J_{i\mu,j\nu}{\bf S}_{i\mu}
\cdot{\bf S}_{j\nu}-\sum_{i\mu}\epsilon_{\mu}
H_{\mu}S^z_{i\mu},\label{eqn_sh}
\end{equation}
where $H_{\mu}$ models the anisotropy field and the collinear
component of the external field on the $\mu$th ion, and the
exchange interaction $J_{i\mu,j\nu}$ depends only on 
$\mu$, $\nu$ and the ionic separation 
${\bf R}_{i\mu}-{\bf R}_{j\nu}$. As usual,
$J_{i\mu,j\nu}=J_{j\nu,i\mu}$ and $J_{i\mu,i\mu}=0$.

The lattice transformation of the exchange constants is a
$n\times n$ Hermitian matrix function and its elements are
\begin{equation}
{\sf J}_{\mu\nu}(\mbox{\boldmath $\kappa$})=
\sum_j J_{j\nu,0\mu}\exp[i\mbox{\boldmath $\kappa$}\cdot
({\bf R}_{j\nu}-{\bf R}_{0\mu})] \label{matrxJ},
\end{equation}
and $\mbox{\boldmath $\kappa$}$ is 
defined within the first Brillouin zone.
Introduce the matrix ${\sf L}(\mbox{\boldmath $\kappa$})$:
\begin{eqnarray}
{\sf L}_{\mu\nu}(\mbox{\boldmath $\kappa$})&\corresponds
&2\Bigl(\epsilon_\mu\sum_{\upsilon}{\sf J}_{\mu\upsilon}(0)
\epsilon_{\upsilon}S_{\upsilon}+\frac{1}{2}H_{\mu}\Bigr)
\delta_{\mu\nu} \nonumber \\
& &-2\epsilon_{\mu}\epsilon_{\nu}(S_{\mu}S_{\nu})^{1/2}
{\sf J}_{\mu\nu}(\mbox{\boldmath $\kappa$}) \label{matrxL}
\end{eqnarray}
which is Hermitian ${\sf L}^{\dag}(\mbox{\boldmath $\kappa$})=
{\sf L}(\mbox{\boldmath $\kappa$})$, and
${\sf L}(-\mbox{\boldmath $\kappa$})
={\sf L}^*(\mbox{\boldmath $\kappa$})$. 
The $n$ eigenvalues, $\lambda_{\mu}(\mbox{\boldmath $\kappa$})$,
of ${\sf E}{\sf L}$ are related to the energies 
of the $n$ spin wave modes by
\begin{equation}
\hbar\omega_{\mu}(\mbox{\boldmath $\kappa$})
=|\lambda_{\mu}(\mbox{\boldmath $\kappa$})|.
\end{equation}
For a stable spin structure, the matrix 
${\sf L}(\mbox{\boldmath $\kappa$})$ is non-negative 
definite for every $\mbox{\boldmath $\kappa$}$, and
$\lambda_{\mu}(\mbox{\boldmath $\kappa$})$ does 
not cross the zero axis for any branch of the dispersion relation, i.e. 
it is either non-negative or non-positive throughout the Brillouin zone.
The number of non-negative (non-positive) branches of
$\lambda_{\mu}(\bbox{\kappa})$ equals the number of
positive (negative) elements of $\epsilon_{\mu}$.

The spin Hamiltonian now can be rewritten in terms of
spin wave modes
\begin{eqnarray}
{\cal H}&=&-N\Bigl(\sum_{j\nu}J_{0\mu,j\nu}S_{\mu}
S_{\nu}+\sum_{\mu}H_{\mu}S_{\mu}\Bigr)\nonumber \\
& &+\hbar\sum_{\mbox{\boldmath $\kappa$},\mu}
\omega_{\mu}(\mbox{\boldmath $\kappa$})
n_{\mu}(\mbox{\boldmath $\kappa$})
\end{eqnarray}
in the semi-classic limit\cite{sw_hp}, namely,
$S_{\mu}-\epsilon_{\mu}\langle S^z_{i\mu}\rangle\ll 2S_{\mu}$.

The primitive magnetic unit cell\cite{bibrew} in the monoclinic AFI
phase of V$_2$O$_3$ is shown in Fig.~\ref{spin_stru}(b) together
with the conventional hexagonal cell of $R\overline{3}c$ symmetry.
Neglecting the small lattice distortion in the phase transition to
the AFI, the monoclinic and hexagonal unit cells are
related by the transformation
\begin{equation}
 \left( \begin{array}{c} {\bf a}_m \\ {\bf b}_m \\ {\bf c}_m
	\end{array} \right) =\left( \begin{array}{rrr}
	1/3 & -1/3 & 1/6 \\
	1/3 & -1/3 & -1/3 \\
	1 & 1 & 0 \end{array} \right) \left(
\begin{array}{c} {\bf a}_h \\ {\bf b}_h \\ {\bf c}_h
	\end{array} \right) 
\end{equation}
and
\begin{equation}
 \left( \begin{array}{c} {\bf a}_h \\ {\bf b}_h \\ {\bf c}_h
	\end{array} \right) =\left( \begin{array}{rrr}
	1 & 1/2 & 1/2 \\
	-1 & -1/2 & 1/2 \\
	2 & -2 & 0 \end{array} \right) \left(
\begin{array}{c} {\bf a}_m \\ {\bf b}_m \\ {\bf c}_m
	\end{array} \right). 
\end{equation}
Denoting the above two matrices as {\sf A} and {\sf B} respectively, 
the corresponding transformation matrices for reciprocal lattice primitive
vectors are {\sf B}$^T$ and {\sf A}$^T$.

There are four magnetic V ions per monoclinic unit cell [refer to
Fig.~\ref{spin_stru}(b)] and
they are assumed to have equivalent spins, $S_{\mu}=S$. 
With the designation of ions, 1-4, in Fig.~\ref{spin_stru}(b),
$(\epsilon_{\mu})=(1,1,-1,-1)$ for the experimentally found
magnetic structure in the AFI. The local field
$(H_{\mu})=(H^+,H^+,H^-,H^-)$
with $H^{\pm} =H_0\pm H_e$, where $H_0$ is the anisotropic field 
in the material and $H_e$ is an external field applied along
the up spin direction.
Limiting the range of exchange interactions  to the 
fourth nearest neighbors, 
there are seven distinct exchange constants: $J_{\alpha},
J_{\beta}, J_{\gamma}, J_{\delta}, J_{\epsilon}, J_{\zeta}$ and
$J_{\eta}$ (refer to Fig.~\ref{spin_stru}(b)). Their contributions
to various spin pairs are tabulated in Table~\ref{tabl_x}.
\begin{table}
\caption{Spin pairs connected by the seven exchange constants.}
\label{tabl_x}
\begin{tabular}{cccl} 
$J_{j\nu,0\mu}$&$\mu$&$\nu$&${\bf R}_{j\nu}-{\bf R}_{0\mu}$\\
\hline
$J_{\alpha}$ & 1 & 2 & {\bf x}\tablenotemark[1]\\
 & 3 & 4 & $-{\bf x}$\\
$J_{\beta}$ & 1 & 3 & ${\bf y}\tablenotemark[2]-{\bf c}_m/2$\\
 & 1 & 3 & ${\bf y}+{\bf c}_m/2$\\
 & 2 & 4 & $-{\bf y}-{\bf c}_m/2$\\
 & 2 & 4 & $-{\bf y}+{\bf c}_m/2$\\
$J_{\gamma}$ & 1 & 2 & ${\bf x}-{\bf a}_m$\\
$J_{\delta}$ & 1 & 3 & ${\bf y}-{\bf a}_m-{\bf c}_m/2$\\
 & 1 & 3 & ${\bf y}-{\bf a}_m+{\bf c}_m/2$\\
 & 2 & 4 & ${\bf a}_m-{\bf y}-{\bf c}_m/2$\\
 & 2 & 4 & ${\bf a}_m-{\bf y}+{\bf c}_m/2$\\
$J_{\epsilon}$ & 1 & 2 & ${\bf x}+{\bf b}_m$\\
$J_{\zeta}$ & 1 & 4 & ${\bf b}_m/2-{\bf c}_m/2$\\
 & 1 & 4 & ${\bf b}_m/2+{\bf c}_m/2$\\
 & 1 & 4 & $-{\bf b}_m/2-{\bf c}_m/2$\\
 & 1 & 4 & $-{\bf b}_m/2+{\bf c}_m/2$\\
 & 2 & 3 & ${\bf b}_m/2-{\bf c}_m/2$\\
 & 2 & 3 & ${\bf b}_m/2+{\bf c}_m/2$\\
 & 2 & 3 & $-{\bf b}_m/2-{\bf c}_m/2$\\
 & 2 & 3 & $-{\bf b}_m/2+{\bf c}_m/2$\\
$J_{\eta}$ & 1 & 1 & ${\bf a}_m$\\
 & 1 & 1\tablenotemark[3] & $-{\bf a}_m$\\
\end{tabular}
\tablenotetext[1]{${\bf x}\equiv {\bf R}_{02}-{\bf R}_{01}=(1/3+2\delta)
({\bf a}_m-{\bf b}_m)$, where $\delta=0.026$.
For ($\mu\nu$)=(21), ${\bf R}_{j\nu}-{\bf R}_{0\mu}
=-{\bf x}$. Exchanging  $\mu$ and $\nu$ switches the sign for 
${\bf R}_{j\nu}-{\bf R}_{0\mu}$.}
\tablenotetext[2]{${\bf y}\equiv(1/3+2\delta)
{\bf a}_m+(1/6-2\delta){\bf b}_m$.}
\tablenotetext[3]{$J_{\eta}$ has identical contributions for 
($\mu\nu$)=(22), (33) and (44) spin pairs.}
\end{table}
Now the matrix {\sf J}($\bbox{\kappa}$) can be readily calculated by 
collecting contributions to spin pairs ($\mu\nu$) according to 
(\ref{matrxJ}).  Notice that spin pairs ($\mu\mu$) are identically 
connected by $J_{\eta}$, therefore
{\sf J}$_{11}$={\sf J}$_{22}$={\sf J}$_{33}$={\sf J}$_{44}$.
Spin pairs (23), (24) and (34) are identically connected as
spin pairs (14), (31) and (21) respectively. This leads to
{\sf J}$_{23}$={\sf J}$_{14}$, {\sf J}$_{24}$={\sf J}$_{31}$
and {\sf J}$_{34}$={\sf J}$_{21}$.
Recall also that {\sf J}($\bbox{\kappa}$) is Hermitian, the
matrix has only four independent elements:
\begin{equation}
 {\sf J}(\bbox{\kappa})=\left( \begin{array}{llll}
A   & B   & C   & D   \\
B^* & A   & D   & C^* \\
C^* & D^* & A   & B^* \\
D^* & C   & B   & A   \end{array} \right). 
\end{equation}
They are given by
\begin{mathletters}
\begin{eqnarray}
A={\sf J}_{11}(\bbox{\kappa})&=&2J_{\eta}\cos(2\pi h),\\
B={\sf J}_{12}(\bbox{\kappa})&=&
J_{\alpha}\exp[i2\pi\phi(h-k)]+\nonumber\\
& & J_{\gamma}\exp[i2\pi(-\psi h-\phi k)]+\nonumber\\
& & J_{\epsilon}\exp[i2\pi(\phi h+\psi k)],\\
C={\sf J}_{13}(\bbox{\kappa})&=&
2J_{\beta}\cos(\pi l)\exp[i2\pi(\phi h+\theta k)]+\nonumber\\
& & 2J_{\delta}\cos(\pi l)\exp[i2\pi(-\psi h+\theta k)],\\
D={\sf J}_{14}(\bbox{\kappa})&=&
4J_{\zeta}\cos(\pi l)\cos(\pi k),
\end{eqnarray}
\end{mathletters}
where $\bbox{\kappa}=(hkl)$ is indexed in the monoclinic reciprocal 
lattice, $\phi=2\delta+1/3$, $\psi=2/3-2\delta$, $\theta=1/6-2\delta$
and $\delta=0.026$.
Notice that only $B$ and $C$ are complex.
$A$ and $D$ are real.

The diagonal terms of {\sf L} in the first line of (\ref{matrxL}) are
\[\delta_{\mu\nu}[H_{\mu}+2S(J_{\alpha}+J_{\gamma}
+J_{\epsilon}+2J_{\eta}-2J_{\beta}-2J_{\delta}-4J_{\zeta})].\]
When there is no external field, $H_{\mu}=H_0$ no longer
depends on $\mu$. In this case,
\begin{equation}
 {\sf L}(\bbox{\kappa})=2S\left( \begin{array}{cccc}
F    & -B  & C  & D  \\
-B^* & F   & D  & C^*  \\
C^*  & D   & F  & -B^* \\
D    & C   & -B & F  \end{array} \right), 
\end{equation}
where 
\begin{equation}
 F=J_{\alpha}+J_{\gamma} +J_{\epsilon}+2J_{\eta}
-2J_{\beta}-2J_{\delta}-4J_{\zeta}+H_0/2S-A.
\end{equation}
The eigen value equation for {\sf EL} is 
\begin{equation}
 \left| \begin{array}{cccc}
F-\lambda'    & -B  & C  & D  \\
-B^* & F-\lambda'   & D  & C^*  \\
C^*  & D   & F+\lambda'  & -B^* \\
D    & C   & -B & F+\lambda'  \end{array} \right|=0, 
\end{equation}
where $\lambda'\equiv \lambda/2S$. It can be verified that the
linear and cubic terms in $\lambda'$ vanish and therefore the
eigen value equation is  quadratic  in $\lambda'^2$:
\begin{equation}
\lambda'^4-2 b(\bbox{\kappa})\lambda'^2+\left| 
{\sf L}(\bbox{\kappa})\right|=0,
\end{equation}
where both the determinant $\left|{\sf L}(\bbox{\kappa})\right|$
and
\begin{equation}
b(\bbox{\kappa})\equiv F^2+|B|^2-|C|^2-D^2
\end{equation}
are real functions of $\bbox{\kappa}$. The consequence of this 
is that
in the absence of an external magnetic field, there are two
branches of doubly degenerate spin wave modes with a dispersion 
relation
\begin{equation}
\hbar\omega(\bbox{\kappa})=2S[b(\bbox{\kappa})\pm
c(\bbox{\kappa})]^{1/2},
\end{equation}
where
\begin{eqnarray}
c^2(\bbox{\kappa})&\equiv &
b^2(\bbox{\kappa})-\left| {\sf L}(\bbox{\kappa})\right|\nonumber\\
&=&4F^2|B|^2+4|C|^2D^2-2|B|^2|C|^2+\nonumber\\
& & B^2C^{*2}+B^{*2}C^2+4FD(BC^*+B^*C).
\end{eqnarray}


\begin{references}
\bibitem{e_used}
{Constant energy scans at 1.8, 3, 6, 9, 12, 15 and 18 meV were used for the
  1.4K fitting; 1.4, 2.5, 4, 9 and 15 meV for 11K; 1.5, 2.5, 4, 9 and 15 for
  30K; 1.5, 7 and 12 meV for 50K; 4, 7, 10, 13 and 17 meV for 70K; 5, 10, 15,
  20 and 28 meV for 90K; 5, 15, 24 and 28 meV for 110K; 1.5, 7 and 12 meV for
  130K; 5, 10, 15 and 28 for 150K; and 1.8, 3, 6, 9, 12 and 18 meV for 200K}.
\bibitem{notea}
{These constants are related to parameters $\Gamma$ and $\overline{A}$ which
we used in \protect\cite{bao96a} by relation: $\Gamma=C a_2^z k_B$ and
$\overline{A}=a_2^z/a_1$.
}
\bibitem{note1}
{The values of individual exchange constants depend strongly 
  on how many of the seven exchanges are included in 
  theoretical fits to our data and they are
  correlated. Thus, our data, as well as those by Word et
  al.\protect\cite{bibrew}, cannot determine them individually. 
  The sum $J_{\beta}+J_{\delta}+2J_{\zeta}= 45.9-2\times4.76\approx 35$ 
  from [\protect\cite{bibrew}], close to our value. 
  The definition of the anisotropy
  $A_{\mu}$ of [\protect\cite{bibrew}] was not given}.
\end{references}
\end{document}